\documentstyle[amstex,12pt,cite,graphicx,chemtex,lscape]{article}

\begin{document}
\author{Andrei L. Tchougr\'{e}eff  $^{1,2}$ \\
and \\
Richard Dronskowski$^{2}$  \\
$^{1}$ Poncelet Laboratory, Independent University of Moscow,\\ 
Moscow Center for Continous Mathematical Education,\\
Bolshoy Vlasyevskiy Pereulok 11, 119002, Moscow, Russia \\
$^{2}$ Institut f\"ur Anorganische Chemie, RWTH Aachen, \\
 Landoltweg 1, 52056 Aachen, Germany
}

\title{Crystal and electronic structure of the room temperature
organometallic ferrimagnet V(TCNE)$_{2}$. Analysis of numerical DoS and magnetic properties as
related to orbital and spin-Hamiltonian models.}
\date{Dedicated to Prof. I.A. Misurkin on occasion of his 75-th birthday}
\maketitle

\begin{abstract}
We present a detailed analysis of the results of our numerical study of the
crystal and electronic structure of the room temperature organometallic
ferrimagnet of general composition V(TCNE)$_{x}$ with $%
x\approx 2$. The results of the LSDA+$U$ study  show that the experimentally determined structure
complies with the magnetic measurements and thus can serve as a prototype
structure for the entire family of the M(TCNE)$_{2}$ organometallic magnets.
The results of the numerical study and of the magnetic experiments are
interpreted using model Hamiltonians proposed here. This allowed us to obtain estimates of the critical
temperature in three- and two-dimensional regimes and to give
an explanation of the differences in behavior of probably isostructural 
V(TCNE)$_{2}$ and Fe(TCNE)$_{2}$ species. 
\end{abstract}

\maketitle

\initial

\section{Introduction}

The most spectacular room temperature organometallic magnet of composition
V(TCNE)$_{x}\cdot y{\rm solvent}$ (where TCNE -- {\bf 1} -- stands for
tetracyanoethylene -- a well known organic electron acceptor; $x\approx 2$
and $y$ depends on the type of the solvent) 
\begin{figure*}[tbp]
\parbox{.24\textwidth}
{\centering
\upethene{NC}{NC}{CN}{CN}
} 
\parbox{.24\textwidth}
{\centering
\hetiisix{D}{CN}{CN}{Q}{NC}{NC}{D}{D}{N}
} 
\parbox{.24\textwidth}
{\centering
\sixringb{Q}{Q}{Q}{Q}{Q}{Q}{}{}{9}
\put (-289,405) {\chemup{NC}{S}{C}{}{}{S}{CN}}
\put (-289,-275) {\cdown{}{}{C}{S}{NC}{S}{CN}}

} 
\parbox{.24\textwidth}
{\centering
\sixring{CN}{CN}{Q}{NC}{NC}{Q}{S}{S}{C}
} \parbox{.24\textwidth}
{\centering
{\bf 1}
} \parbox{.24\textwidth}
{\centering
{\bf 2}
} \parbox{.24\textwidth}
{\centering
{\bf 3}
} \parbox{.24\textwidth}
{\centering
{\bf 4}
}
\end{figure*}
attracts a lot of attention since the time it had been synthesized yet in
the beginning of the 1990's \cite{Manriquez}. It is an amorphous moisture
sensitive precipitate with the outstanding critical temperature of the
transition to the magnetically ordered state 
estimated to be of {\it ca. }400 K.\footnote{The magnetic momenta are 
spontanelusly predominantly aligned in one direction below the critical temperature.  
} It is higher than the
decomposition temperature ({\it ca. } 350 K ) which singles it out among its
numerous analogs with a variety of involved organic acceptors
(Ref. \cite{Vickers1} -- tetracyanopyrazine -- {\bf 2}; Ref. \cite{Vickers2}
-- 7,7,8,8-tetracyano-p-quinodimethane -- {\bf 3}; Ref. \cite{Taliaferro} --
tetracyanobenzene -- {\bf 4}) and of the metals (Ref. \cite{Pokhodnya} --
iron), synthesized in the following years, since none of them manifested as
fascinating magnetic properties as the very first V(TCNE)$_{2}$ compound (see
Table 1 
\cite{MillerEpstein}).

Generally one has to say that not only the critical temperature, but also
other properties of the compounds of the considered class are sensitive to
the details of the preparation procedure and/or the solvent employed. 
For example, the V-TCNE compound
is known in two forms. The original of Ref. \cite{Manriquez} coming from
the reaction of V(C$_{6}$H$_{6}$)$_{2}$ with TCNE in the CH$_{2}$Cl$_{2}$
solution or  obtained from V(CO)$_{6}$ with use of the CVD technique 
exhibits the saturation magnetization at the zero temperature 
 which corresponds to approximately one unpaired electron per formula unit 
and in fact is somewhat lower than this value. However in Ref. \cite{ZhouLongMillerEpstein} 
(see also a review 
\cite{MillerEpstein}) a V(TCNE)$_{2}$ compound was reported containing tetrahydrofurane
as a solvent with the saturation magnetization 
almost twice as strong compared to that of the original compound.
For that reason hereinafter we shall refer to these two forms of the V(TCNE)$_{2}$ compound
as the highly magnetic (HM) and the low magnetic(LM) ones. 

For more than one decade the amorphousity of the compound of interest did
not allow anyone to make whatever definitive conclusion concerning its
structure. The critical breakthrough became possible with the recent 
work \cite{Her} where the authors were able to establish the structure of
the Fe$^{2+}$ analog (presented in Fig. 1
) 
of the V(TCNE)$_{2}$ compound 
using Rietveld refinement of the synchrotron powder
diffraction data
and to reveal its most remarkable features:
the presence of the dimer form of the TCNE$^{\dot{-}}$ radical-anion: $\left[
\text{TCNE}\right] _{2}^{2-}\ $= C$_{4}$(CN)$_{8}^{2-}$ playing an important
role in shaping the loose three-dimensional structure and assuring as well
the three-dimensional character of magnetic interactions in the system.

In our previous paper Ref. \cite{Tch082} we were able to demonstrate that the
experimental structure represented in Fig. 1 
can be easily
related with one represented in Fig. 2a 
proposed yet in Ref. \cite%
{Tchougreeff-Hoffmann} 
in order to conform with the magnetic data on the V-TCNE\ compound
available that time. That latter represented a simple cubic lattice with the vertices
occupied by the vanadium ions. Two $ab$-faces of each cube 
 remain empty whereas four others are filled
by the TCNE units forming channels extended in the $c$-direction. In the
structure presented in Fig. 2a 
the C=C bonds of the TCNE units are located in the same
plane (as well perpendicular to the $c$-axis of the lattice) so that they
are orthogonal to each other. It is not the unique possibility.
An alternative structure differing from that of
Fig. 2a 
by rotating the TCNE units placed in the $bc$-faces
 by 90$^{\circ }$ in their respective planes is also possible.
The result of such a rotation is presented in Fig. 2b
. 
In this
structure the central C=C bonds of the TCNE units are as previously
orthogonal, but now they lay in orthogonal planes so that the axes
of these bonds do not intersect rather cross each other. This structure has
been called the "principal" structure in Ref. \cite{Tch082}. The principal
structure Fig. 2b 
can be easily put in the relation
with the experimental one. If in the quadrupled unit cell $2a,2b,c$ of the
principal structure four TCNE units extended in the $b$-direction are
allowed to pairwisely rotate towards each other so that single C--C bonds can form
between respective ethylenic carbon atoms thus yielding the [TCNE]$_{2}^{2-}$%
=C$_{4}$(CN)$_{8}^{2-}$ dimers and the V-TCNE sheets originally laying in the $ac$-planes
are accordingly ruffled one finally arrives to the experimental
structure of Fe(TCNE)$_{2}$ (Fig. 1
). Intermediate structures
along this hypothetical reaction path are presented in Fig. 3
. 
For this sequence of structures we performed in Ref. \cite{Tch082} the LSDA + 
$U$ calculations with use of the VASP program suite Ref. \cite{VASP}
 of the respective electronic structures and energies. Specifically,
the PAW potential has been used and the values of the $U$ parameter for the V, N, and C 
atoms were respectively taken as 
5.344, 4.840, and 4.428 eV. 
 The
calculations have been performed as follows: first the principal structure
with the equalized lattice parameters have been quadrupled. This structure has been
connected with the experimental structure by a straight line in the
configuration space. Further details of the numerical treatment can be found
in Ref. \cite{Tch082}. Its results are reproduced in Fig. 4
.

The overall densities of states in two spin channels as given in Fig. 4 
are obviously difficult to understand. One can only see some
spin polarization of the upper filled bands as well as an evolution of a
noticeable density of states near the Fermi level present in both spin
channels in the initial state to the final state where the expected
spin-polarized structure can be recognized which does not show any DoS in
either of the spin channels at the Fermi level.  
In this paper we present a detailed analysis of the results of our
numerical studies Ref. \cite{Tch082}\ of the models of V-TCNE room
temperature organometallic ferrimagnet and expressed them in terms of
the effective spin-Hamiltonian for a selection of interacting
atomic/molecular states. The proposed model is then applied to analysis of a
wider collection of experimental data available for this fascinating object
and its Fe analog.

\section{Detailed analysis of DoS}

In order to analyse the details of the evolution of the DoS features along
the path going from quadrupled principal structure to the
experimental one we performed a detailed study of projections of the DoS.
In the initial structure the three bands in the 
$- 9 \div - 11 $ eV range are predominantly C-bands. Other 
bands are the well hybridized C-N-bands except the "spin-left" band right below the 
Fermi level, which is predominantly contributed by the vanadium states. As one can see
from Fig. 6 
these are the $d$-states of vanadium atoms. Some vanadium stemming DoS
in the range $- 8 \div - 10 $ eV are the $s$- and $p$-states of vanadium involved in the bonding with the
nitrogen atoms. This can be seen on the corresponding N-projection of DoS
in Fig. 7 
from which one can deduce that the $s$- and $p$-states of vanadium hybridize
predominantly with the N-states coming from the V-TCNE layers (see below). 
In order to deepen our understanding of the DoS presented in Fig. 5 
we notice that even in that complex system the atoms
involved are relatively easily classified into types. One can distinguish
transition metal ions whose $d$-density is expected to contribute
significantly to the spin polarized bands, the nitrogen atoms where one can
expect significant DoS changes due to expected break of the pairs of excessive V-N
bonds while going from the quadrupled principal structure to the
experimental one. Analogously along the same route one can expect
remarkable variations in the projection of the DoS to the carbon atoms forming the
C-C bonds in the [TCNE]$_{2}^{2-}$ units and to the carbon and nitrogen
atoms in the ruffled V-TCNE layers. Other features of the DoS, however, are not
expected to significantly vary along the path. Considering the evolution of
the above-mentioned projections of the DoS along the path Fig. 3 
as represented in Figs. 4 
- 6 
one can see the expected
reconstruction of the projected DoS's. For example,  almost
nothing happens to the $d$-bands along this path. They remain rigid
triply degenerate ones and do not change
their position relative to the Fermi level despite the fact that the
coordination number of vanadium ions changes from eight in the
quadrupled principal structure to six in the experimental one. This agrees
with the numerical result concerning the distribution of spin polarization
in the direct space: for all structures depicted in Fig. 3 
the magnetic moments residing in the $d$%
-shells of vanadium ions range from 2.615 for the first (quadrupled
principal)\ structure to 2.582 for the last (experimental) one {\it i.e.}
are almost constant. The significant variation of the total magnetic moment along the
"reaction path" observed in our numerical experiment is to be almost
completely attributed to that of the magnetic moments residing in the
"organic" part of the organometallic magnet. This is precisely what one
should expect within the general picture including the formation of the [TCNE]$%
_{2}^{2-}$ dimers. On the other hand the stability of the $d$-bands along
the path can be only understood if one assumes that the break of two V-N bonds at
each atom is at least partially compensated by shortening (strengthening) of
other two bonds extended in the $b$-direction.

By contrast the projections of DoS to the characteristic organogenic atoms
significantly modify along the \ "reaction path". 
For example, the
dominant part of the N-DoS is
concentrated in two broad bands at {\it ca.} $-5$ and {\it ca.} $-6 \div -6.5$ eV. The 
lower of the two is almost equally contributed by the nitrogen atoms
of all three types and its position is stable throughout the path. 
The same applies to somewhat smaller contribution from the bonding N-atoms
extended in the $b$-direction to this band. Their contribution to the upper 
of the two mentioned bands is noticeably spin-polarized. The density
from the N atoms bearing the dangling lone pairs fairly manifests itself 
in the same band. Weak features in the N-DoS can be 
observed in the 4 eV wide range right below the Fermi level. The peak 
right below the Fermi level
in the
"right-spin" channel in the initial structure is equally contributed by the 
"bonding",  "to be dangling", and "magnetic" nitrogens. (In general
the DoS projections for the "bonding" and  "to be dangling" nitrogens 
coincide in the initial structure). In the final structure this peak 
splits so that its upper part ("right-spin" channel right below the Fermi level)
is contributed by the "magnetic" nitrogens, whereas two peaks next to it in the bottom
direction is contributed by the "bonding" nitrogens. The contribution from
the "dangling" nitrogens contributes to the upper part of the wide band at {\it ca.} $-3.5 \div -5.5$ eV. 

Of particular interest is the evolution of the C-projection of the DoS.
That on the ethylenic carbon atoms is most interesting. The ethylenic carbons can
be further subdivided in two categories: those in the planes extended in the 
$a$ and $c$ horizontal directions and those extended in the vertical ($b$)
direction and further involved in the rotation yielding the
[TCNE]$_{2}^{2-}$ units. At the initial stage both projected DoS are
significantly polarized and both noticeably contribute to the "right-spin" density right below the
Fermi level (the DoS projection to the to be bonding atoms 
is not seen in the leftmost graph of Fig. 6 since at this point they are 
degenerate with the "magnetic" projection and are masked by these latter).
 From further graphs of the DoS projected to the atoms forming
the emerging C-C bonds one can conclude that such bonds represented in our study by 
two spin sub-bands for the left and right spin channels completely develop at a pretty late stage of
the hypothetical transition: in the middle of the path the corresponding peaks in the 
density projections can be yet clearly seen in both spin channels near to the 
spin-polarized DoS of the "magnetic" atoms, although the peaks corresponding to
 the emerging bonds are not polarized. 
 By contrast the projection of DoS on the magnetic
C-atoms in the horizontal planes develop two sub bands right above and below
the Fermi level of which the lower one (right-spin) is completely occupied by electrons
with the spin projection opposite to that of the electrons occupying the $d$%
-subband. This is in the fair agreement with the assumption concerning the
nature of the subbands located in the vicinity of the Fermi level made in
our previous paper Ref. \cite{Tch082}. 

\section{Model Hamiltonians for V-TCNE system }

\subsection{Model Orbital Hamiltonian}

Fascinating properties of the V(TCNE)$_{2}$ magnet call not only for
numerical modelling, but also for some qualitative picture. However, for the final
("experimental") structure all the DoS projections manifest themselves as
very narrow bands. This indirectly indicates that the band picture used throughout the
calculations is
not completely adequate and that an adequate model must be given in terms
of an effective Hamiltonian representing the electronic structure of the HM
V(TCNE)$_{2}$ magnet (in its "experimental" structure) in terms of some objects
local in the direct space {\it e.g.} local spins similar to that proposed
yet in Ref. \cite{Tchougreeff-Hoffmann}. As in the case of band models the
most important one-electron states to be included are these contributing to
the energy bands in the vicinity of the Fermi level. Based on our analysis
of projected DoS performed in the previous Section we can conclude that the
states of the [TCNE]$_{2}^{2-}$ units contribute to the bands far away from 
the Fermi level. The DoS related to this unit goes away from the
Fermi level along the "reaction path". Thus the observed electronic
structure is primary one of  the individual (ruffled) V-TCNE layer extended
in the $ac$-plane. For constructing the Hamiltonian for this layer one can
employ the unit cell of the principal model dropping from it the TCNE unit
extended in the $b$-direction (and finally engaged in formation of the [TCNE]%
$_{2}^{2-}$= C$_{4}$(CN)$_{8}^{2-}$ dimers).
According to Ref. \cite{Tchougreeff-Hoffmann} on the vanadium sites it
suffice to consider only the $d$-shells of the metal ions which is also confirmed
by our current analysis. The overlap of
the $d$-shell of the metal ions with the $\sigma $-orbitals of the TCNE's
(including those implied in the model) ensures the standard two-over-three
splitting of the $d$-shell characteristic for the octahedral environment. In
the case of vanadium three unpaired electrons in the $d$-shell occupy
respectively three orbitals in the $t_{2g}$-manifold.\ The basis orbitals $%
d_{xy}$, $d_{xz}$, and $d_{yz}$ can be characterized by the normal to the
plane in which each of the orbitals lays -- $\zeta $, $\eta $, and $\xi $ --
are subsequently used in the notation. For the donor sites in the
 ruffled planes the $b_{3g}\left( \pi ^{\ast }\right) $ LUMOs of
the TCNE (singly occupied in the radical anion) are included. In each such a
layer each metal ion is surrounded (coordinated) by four TCNE units which
are in their turn coordinated to (surrounded by) four metal atoms. In this
case the layer unit cell composition V:TCNE is 1:1.

The model Hamiltonian for the V(TCNE)$_{2}$ magnet, formulating the above
ideas, has the general form: 
\begin{equation}
H=\sum_{{\bf r}}\left( H_{d}({\bf r})+H_{a}({\bf r})+H_{da}({\bf r})+H_{dd}(%
{\bf r})\right)   \label{Hamiltonian}
\end{equation}%
The contributions to it are the following. Operator $H_{a}({\bf r})$
describes electrons in the acceptor orbital of the TCNE$^{\dot{-}}$
radical-anion in the ${\bf r}$-th unit cell: 
\begin{equation}
\label{a-Hamiltonian}
\begin{array}{rcl}
H_{a}({\bf r}) &=&-\alpha _{a}\hat{n}_{a{\bf r}}+U_{aa}\hat{n}_{a{\bf r}\downarrow }
\hat{n}_{a{\bf r}\uparrow }  \\ 
\hat{n}_{a{\bf r}\sigma } &=&a_{{\bf r}\sigma }^{+}a_{{\bf r}\sigma };
\hat{n}_{a{\bf r}}=\sum_{\sigma }\hat{n}_{a{\bf r}\sigma }  
\end{array}
\end{equation}%
Symbol $a_{{\bf r}\sigma }^{+}(a_{{\bf r}\sigma })$ is the operator
creating (annihilating) an electron with the spin projection $\sigma $ on the
acceptor orbital of the TCNE molecules in the ${\bf r}$-th unit cell. In
eq. (\ref{a-Hamiltonian}) the first term is the energy of attraction of an
electron to the core of TCNE -- the orbital energy of the $b_{3g}\left( \pi
^{\ast }\right) $ LUMO shifted by the electrostatic field induced by the
entire crystal environment. The second term in  eq. (\ref%
{a-Hamiltonian}) is the Hubbard one, effectively describing the Coulomb
repulsion of electrons with opposite spin projections eventually occupying
the same acceptor orbital.

The operator $H_{d}({\bf r})$ describes electrons in the $t_{2g}$-subshell
of the $d$-shell of the vanadium ion in the ${\bf r}$-th unit cell: 
\begin{eqnarray}
H_{d}({\bf r}) &=&\left[ -\alpha _{d}(\hat{n}_{\zeta {\bf r}}+\hat{n}_{\eta 
{\bf r}}+\hat{n}_{\xi {\bf r}})\right. +  \nonumber \\
&&+\left. (U_{dd}+2J_{dd})(\hat{n}_{\zeta {\bf r}\downarrow }\hat{n}_{\zeta 
{\bf r}\uparrow }+\hat{n}_{\eta {\bf r}\downarrow }\hat{n}_{\eta {\bf r}%
\uparrow }+\hat{n}_{\xi {\bf r}\downarrow }\hat{n}_{\xi {\bf r}\uparrow })%
\right] +  \nonumber \\
&&+\frac{(U_{dd}+J_{dd}/2)}{2}\sum_{\sigma ,\sigma ^{\prime }}(\hat{n}%
_{\zeta {\bf r}\sigma }\hat{n}_{\xi {\bf r}\sigma {\bf ^{\prime }}}+\hat{n}%
_{\zeta {\bf r}\sigma }\hat{n}_{\eta {\bf r}\sigma {\bf ^{\prime }}}+\hat{n}%
_{\xi {\bf r}\sigma }\hat{n}_{\eta {\bf r}\sigma {\bf ^{\prime }}})+
\label{d-Hamiltonian} \\
&&-4J_{dd}(\hat{S}_{\zeta {\bf r}}\hat{S}_{\xi {\bf r}}+\hat{S}_{\zeta {\bf r%
}}\hat{S}_{\eta {\bf r}}+\hat{S}_{\xi {\bf r}}\hat{S}_{\eta {\bf r}}) 
\nonumber \\
\hat{n}_{\gamma {\bf r}\sigma } &=&\gamma _{{\bf r}\sigma }^{+}\gamma _{{\bf %
r}\sigma },\ \hat{n}_{\gamma {\bf r}}=\sum_{\sigma }\hat{n}_{\gamma {\bf r}%
\sigma };\gamma =\xi ,\ \eta ,\zeta .  \nonumber
\end{eqnarray}%
In eq. (\ref{d-Hamiltonian}) $ \hat{n}_{\gamma {\bf r}%
\sigma }$ are the operators of the number 
of electrons with the spin projection $\sigma $ on the $d_{xy},d_{yz},$ and $%
d_{xz}$ orbitals of the vanadium ion in the ${\bf r}$-th unit cell. The spin operators and spin-operator
product terms are defined by the well-known relations: 
\begin{eqnarray*}
\hat{S}_{\gamma {\bf r}}\hat{S}_{\gamma ^{\prime }{\bf r}} &=&1/2(\hat{S}_{\gamma {\bf r}}^{+}%
\hat{S}_{\gamma ^{\prime } {\bf r}}^{-}+\hat{S}_{\gamma ^{\prime }{\bf r}}^{+}\hat{S}_{\gamma {\bf r}}^{-})+%
\hat{S}_{\gamma {\bf r}}^{z}\hat{S}_{\gamma ^{\prime } {\bf r}}^{z} \\
\hat{S}_{\gamma {\bf r}}^{+} &=&\gamma _{{\bf r}\uparrow }^{+}\gamma _{{\bf r}\downarrow },\hat{S}%
_{\gamma {\bf r}}^{-}=\gamma _{{\bf r}\downarrow }^{+}\gamma _{{\bf r}\uparrow },\hat{S}_{\gamma
{\bf r}}^{z}=1/2(\hat{n}_{\gamma {\bf r} \uparrow }-\hat{n}_{\gamma {\bf r} \downarrow }).
\end{eqnarray*}
where the symbols $\gamma _{{\bf r}\sigma }^{+}\
(\gamma _{{\bf r}\sigma })$ represent 
the operators creating (annihilating)
an electron with the spin projection $\sigma $ on the $d_{xy},d_{yz},$ and $%
d_{xz}$ orbitals of the vanadium ion in the ${\bf r}$-th unit cell. 

The
first row in the above operator describes the attraction of electrons in the 
$d$-orbitals to the cores of vanadium ions (shifted by the electrostatic
field of the rest of the crystal). Two further rows describe the
spin-symmetric part of the Coulomb interaction of electrons in the $d$%
-shell. The last row describes the spin dependent part of the Coulomb
interaction of electrons in the $d$-shell (exchange). It is ultimately
responsible for the Hund's rule in atoms and for the high spin of the ground
state of electrons in the $d$-shell.
The contributions to the Hamiltonian eq. (\ref{Hamiltonian}) described so
far model isolated local states important for the crystal description. The
magnetic order can only be possible due to various interaction terms.
Operator $H_{da}({\bf r})$ describes the electron hopping between the $d$%
-states of vanadium ions and the acceptor states. The $d_{xy}$-state
represented by the $\zeta _{{\bf r}\sigma }^{+}\ (\zeta _{{\bf r}\sigma })$
operators being of the (approximate) $\sigma $-symmetry with respect to the $%
ac$ plane (the ruffling of the V-TCNE plane is neglected) has no overlap
with the LUMO's of TCNE's which are (again approximately) of the $\pi $%
-symmetry with respect to the same plane. Two others ($d_{xz}$- and $d_{yz}$%
-states represented respectively by the $\eta _{{\bf r}\sigma }^{+}\ (\eta _{%
{\bf r}\sigma })$ and $\xi _{{\bf r}\sigma }^{+}\ (\xi _{{\bf r}\sigma })$
operators) overlap with the LUMOs of two (different) neighbor TCNE units
each. The phase relations between the orbitals involved in the model lead to
such a distribution of signs at the one-electron hopping parameters that the 
hopping operator acquires the form: 
\begin{equation}
H_{da}({\bf r})=-t_{da}\sum_{\sigma }{}\left[ \xi _{{\bf r}\sigma
}^{+}\left( a_{{\bf r}\sigma }+a_{{\bf r}+{\bf a}+{\bf c}\sigma }\right)
-\eta _{{\bf r}\sigma }^{+}\left( a_{{\bf r}+{\bf a}\sigma }+a_{{\bf r}+{\bf %
c}\sigma }\right) \right] +h.c.  \label{da-interaction}
\end{equation}%
where the parameter $t_{da}>0$ describes the magnitude of the hopping
between the acceptor state and the neighbor $d$-state.

The sum of the above contributions to the effective Hamiltonian in fact form
that for an isolated V-TCNE layer. In the "experimental" structure the
diamagnetic C$_{4}$(CN)$_{8}^{2-}$ units seem to effectively isolate the
V(TCNE) sheets from each other. Nevertheless, one should assume that certain
indirect interaction between the $d$-states in the $b$-direction is possible
through the mediation of the [TCNE]$_{2}^{2-}$ units. It was proposed in
Ref. \cite{Tch082} to use an effective hopping similar to eq. (\ref%
{da-interaction}). Since it is any way an effective interaction it can be
chosen in a way which fits better to the method the system is treated. For
this reason we postpone the discussion of this term. 

In our previous paper Ref. \cite{Tch082} we considered the band
model of the V-TCNE organometallic magnet as derived from the orbital
Hamiltonian eqs. (\ref{Hamiltonian}) - (\ref{da-interaction}) and employed
them for analysis of results of our numerical experiments performed with use
of the VASP package. These latter are, however, in a kind of fundamental
contradiction with the physics of the system at hand. This manifests itself
in the very narrow bands coming out of calculation, as we already mentioned.
The reason is that the hopping parameter $t_{da}$ entering eq. (\ref{da-interaction}) which are generally responsible
for extension of one-electron states over the crystal (band formation) and
which are proportional to the overlap between the orbitals represent the smallest
energy scale in the system. Generally it leads to a break of the delocalized
(band) picture and makes a local description to be more adequate. The
latter can be sequentially derived by treating perturbatively
 the hopping operator eq. (\ref{da-interaction}). It
yields the effective Hamiltonian of the Heisenberg form in terms of the
spins of electrons occupying the local states (orbitals) involved. Its parameters
are estimated in Appendix \ref{SpinHamiltonian}.

The overall result comes out as
a spin Hamiltonian of the form: 
\begin{equation}
\begin{array}{ccc}
H_{{\rm spin}}^{{\rm layer}} & = & -4J_{dd}\sum\limits_{{\bf r}}(\hat{S}_{\zeta {\bf r}}%
\hat{S}_{\xi {\bf r}}+\hat{S}_{\zeta {\bf r}}\hat{S}_{\eta {\bf r}}+\hat{S}%
_{\xi {\bf r}}\hat{S}_{\eta {\bf r}})+ \\ 
&  & +2K_{da}\sum\limits_{{\bf r}}\left[ \hat{S}_{\xi {\bf r}}\left( \hat{S}%
_{a{\bf r}}+\hat{S}_{a{\bf r}+{\bf a}+{\bf c}}\right) +\hat{S}_{\eta {\bf r}%
}\left( \hat{S}_{a{\bf r}+{\bf a}}+\hat{S}_{a{\bf r}+{\bf c}}\right) \right] 
\end{array}
\label{spin-Hamiltonian}
\end{equation}%
which describes effective magnetic interactions in an isolated layer. It
must be complemented by interlayer interactions. If the vanadium ions in
adjacent layers\ are coupled by an an effective hopping an analogous
perturbative procedure results in the antiferromagnetic sign of the
effective magnetic interaction. This contradicts to the existence of the
nonzero overall magnetization in the V-TCNE\ magnets below the critical
temperature (with the antiferromagnetic interlayer coupling the
magnetization of one layer would be cancelled by that of another). For that
reason we have to supplement the Hamiltonian eq. (\ref{spin-Hamiltonian}) by
an effective interlayer interaction with the ferromagnetic sign of the
corresponding exchange parameter. It cannot directly come from any
perturbative treatment of the hopping. By contrast some mechanism of
ferromagnetic coupling described {\it e.g.} in Refs. \cite{Goodenough} or 
\cite{Kahn} and implemented in papers \cite{Tch018,Tch019}
devoted to the exchange in metallocene based organometallic magnets
(Miller-Epstein magnets) acting through the [TCNE]$_{2}^{2-}$ units might be
expected. Indeed as one can see from the Fig. 7 
despite the
fact the the states located in the [TCNE]$_{2}^{2-}$ units are pulled up and
down from the vicinity of the Fermi level, some spin
polarization of these bands particularly of those which are contributed by
the "bonding" nitrogens indicates the involvement of the [TCNE]%
$_{2}^{2-}$ units in transfer of magnetic interactions between the $d$%
-shells in the $b$-direction (between the layers). 

With this {\it caveat} the spin Hamiltonian written in terms of "true"
electronic spins is the following:%
\begin{equation}
\begin{array}{ccc}
H_{{\rm spin}} & = & -4J_{dd}\sum\limits_{{\bf r}}(\hat{S}_{\zeta {\bf r}}%
\hat{S}_{\xi {\bf r}}+\hat{S}_{\zeta {\bf r}}\hat{S}_{\eta {\bf r}}+\hat{S}%
_{\xi {\bf r}}\hat{S}_{\eta {\bf r}})+ \\ 
& + & 2K_{da}\sum\limits_{{\bf r}}\left[ \hat{S}_{\xi {\bf r}}\left( \hat{S}%
_{a{\bf r}}+\hat{S}_{a{\bf r}+{\bf a}+{\bf c}}\right) +\hat{S}_{\eta {\bf r}%
}\left( \hat{S}_{a{\bf r}+{\bf a}}+\hat{S}_{a{\bf r}+{\bf c}}\right) \right]
\\ 
& + & 2K_{dd}\sum\limits_{{\bf r}}\left( \hat{S}_{\xi {\bf r}}+\hat{S}_{\eta 
{\bf r}}+\hat{S}_{\zeta {\bf r}}\right) \left( \hat{S}_{\xi {\bf r+b}}+\hat{S%
}_{\eta {\bf r+b}}+\hat{S}_{\zeta {\bf r+b}}\right)%
\end{array}
\label{true-spin-Hamiltonian}
\end{equation}
The Hamiltonian eq. (\ref{true-spin-Hamiltonian}) is not a standard
Heisenberg  Hamiltonian  usually used to describe magnetic properties
of insulators. This latter is written in terms of the effective local
spins residing at each atomic magnetic center. In our case the  vanadium ions 
represetn such nontrivial magnetic centers bearing
effective spins in the $d$-shells: 
\begin{equation}
\hat{S}_{d{\bf r}}=\sum_{\gamma }\hat{S}_{\gamma {\bf r}}.
\label{effective-spins}
\end{equation}%
According to Ref. \cite{Vonsovsky} using the effective spins is, however, an
approximation since the transition from the representation of the effective
Hamiltonian in terms of the of individual electronic spins-$\frac{1}{2}$  eq. (%
\ref{true-spin-Hamiltonian}) which can be sequentially derived from the
model orbital Hamiltonian eqs. (\ref{Hamiltonian}) - (\ref{da-interaction})
by perturbative treatment of the hopping term eq. (\ref{da-interaction}) to
the phenomenological Hamiltonian eq. (\ref{phen-Hamiltonian}) operating with
the effective spins eq. (\ref{effective-spins}) is only possible if the
exchange interactions of all individual spins in one magnetic center (in our
case -- the V ion) with those in the other magnetic center (in our case the
effective spin in TCNE$^{\dot{-}}$ coincides with the individual one) are
equal. This is obviously not the case since the electronic spin $\hat{S}%
_{\zeta {\bf r}}$ in the $d$-shell to the first approximation does not
interact with the spin residing in  any of acceptor orbitals.

This generally poses the problem since in the Hamiltonian eq. (\ref%
{true-spin-Hamiltonian}) at least the exchange parameters $J_{dd}$ and $K_{da}
$ can be independently determined respectively by eq. (\ref{Heisenberg-exchange-parameter}%
) and atomic spectra, whereas the parameter $J_{\bot }$ in eq. (\ref{phen-Hamiltonian}) remains
completely empirical quantity. The fact that $J_{dd}\gg K_{da}$ in eq. (\ref%
{true-spin-Hamiltonian}) allows to approximately replace the spins of
separate electrons in the ${\bf r}$-th $d$-shell by the operator of the
total spin of the respective $d$-shell. Further details of this transition
are given in Appendix \ref{SpinToPhen}.

The phenomenological Hamiltonian written in terms of the effective spins eq.
(\ref{effective-spins}) to be used for modeling the entire crystal is:%
\begin{equation}
\begin{array}{ccc}
H_{{\rm phen}} & = & J_{\Vert }\sum\limits_{{\bf r}}\hat{S}_{d{\bf r}}\left[
\hat{S}_{a{\bf r}}+\hat{S}_{a{\bf r}+{\bf a}}+\hat{S}_{a{\bf r}+{\bf c}}+%
\hat{S}_{a{\bf r}+{\bf a}+{\bf c}}\right] + \\ 
&  & +J_{\bot }\sum\limits_{{\bf r}}\hat{S}_{d{\bf r}}\hat{S}_{d{\bf r}+%
{\bf b}}%
\end{array}
\label{phen-Hamiltonian}
\end{equation}%
In the next Section we apply it to analysis of magnetic properties of the 
HM V(TCNE)$_{2}$ material.

\section{Magnetic properties of V(TCNE)$_{2}$ as interpreted with use of
 phenomenological Hamiltonian}

The magnetic and thermodynamic properties of the HM V(TCNE)$_{2}$ material must be 
derived from the phenomenological Hamiltonian  eq. (\ref{phen-Hamiltonian}). 
Two types of data will be of interest for us: the critical temperature 
of transition into magnetically ordered state and the temperature dependence 
of spontaneous magnetization. The mean field estimates of the critical temperature
 used so far in the literature 
lack the account of structural information. It is important to realize that the quantities
of interest are sensitive to these details as represented in the respective Hamiltonians. 

The commonly used (see \emph{e.g.} Ref. \cite{MillerEpstein}) 
symmetric mean field formula:
\[
\theta_{N}^{MF}=\frac{zJ_{eff}}{3}S_{d}\left(S_{d}+1\right)
\]
(where $T=\theta/k_{B}$) ignores the acceptor spins and sets $z$ to be the number of indirectly neighboring magnetic
metal ions. It yields 
the quadratic dependence of the critical temperature on the spin
of the metal ion. 
On the other hand according to Ref. \cite{Her} the
critical temperature for the magnetically ordered state of a material with
two types of spins ($S_{d}$ and $S_{a}(=1/2)$) is described by 
the mean field formula: 
\[
\theta^{\rm MF}_{N}=\frac{\left\vert J^{\rm MF}_{{\rm eff}}\right\vert}{3} \sqrt{Z_{ad}Z_{da}%
}\sqrt{S_{d}(S_{d}+1)S_{a}(S_{a}+1)} 
\]%
(The factor of two is dropped here to get the formula to conform with the 
Hamiltonian definition accepted in the present paper). This formula appears as a 
zero interlayer coupling limit of the  mean field expression for the N\'{e}el
temperature in a {\em ferrimagnet} eq. (\ref{MeanFieldTheta}) derived in Appendix \ref{MeanFieldSection}. 
In Ref. \cite{Her} it had been applied to the Fe(TCNE)$_{2}$ compound for which the
structure measurements have been performed there. It, however, brings up two
complications -- one theoretical and another experimental. From the experimental point of view we notice 
that the above formula as well as formula eq. (\ref{MeanFieldTheta}) is effectively linear in $S_{d}$
rather than quadratic (in the high anisotropy limit). For that reason even the mean field estimates of the
exchange parameters as given in Table 1 must be reconsidered since the latter 
had been obtained with use of the quadratic dependence. As one can see
from the structure the choice of $Z_{ad}=Z_{da}=4$ yields the 
estimate of the mean field exchange parameter for the Fe(TCNE)$_{2}$ compound 
 of $J^{\rm MF}_{{\rm eff}}({\rm Fe})= $
43 K and for the V(TCNE)$_{2}$ compound -- $J^{\rm MF}_{{\rm eff}}({\rm V})= 183$ K. Both values are significantly
larger than those given in Table 1, but it is remarkable that the difference 
between them (to be explained) reduces from the factor of larger than five to that of 4.25. 
We notice that following the assumption of Ref. \cite{MillerPreprint} and using 
$Z_{dd}=6$ for the V(TCNE)$_{2}$ compound further reduces the difference and the value
of the exchange parameter for the latter, but from our point of view it cannot be
substantiated within the scope of the model considered in the present paper.    

From the theoretical point of view, even the improved molecular field expression eq. (\ref{MeanFieldTheta}) has that disadvantage that 
it predicts a nonvanishing ordering temperature for $ J_{\perp } = 0$.
It is obviously wrong and such an estimate is not acceptable in the context where a strong anisotropy might be expected
on the structure basis.
As it has been mentioned the model must be complemented
by the interlayer interactions between the effective spins $3/2$ 
on the vanadium sites mediated by the
diamagnetic (closed shell) [TCNE]$^{2-}$ units. Remarkably enough
the sign of this interaction must be {\em ferromagnetic} (the magnetic
moments residing in the layers must be pointing in the same direction) to
ensure the existence of the net spontaneous magnetization in the
three-dimensional sample, although in general one has to expect {\em %
antiferromagnetic} sign of such an interaction \cite{Goodenough} (see below). 

In a presumably rather
anisotropic situation brought by tentative difference in mechanisms of the
intralayer and interlayer interactions (respectively "antiferromagnetic
kinetic exchange" for the
intralayer interaction and the "ferromagnetic superexchange" for the interlayer
one) the critical temperature has to be estimated from the spin-wave treatment
 taking an adequate care about the anisotropy of the effective
spin-spin interaction and at least providing
a correct asymptotic value of the critical temperature for the vanishing interlayer coupling $%
J_{\perp }$.
This is done by the formula 
\begin{equation}
M_{s}=M_{0}\left[1-\left(\frac{\theta}{\theta_{\mathrm{N}}}\right)^{\frac{3}{2}}\right]\label{BlochMagnetization}\end{equation}
expressing the Bloch $T^{\frac{3}{2}}$ law for the 
temperature dependence of the spontaneous magnetization as derived in Appendix \ref{SpinWave} with 
the critical (N{\'e}el) temperature given by:
\begin{equation}
\theta_{\mathrm{N}}=\frac{4\pi}{\zeta^{\frac{2}{3}}(\frac{3}{2})}\frac{1}{S_{d}-S_{a}}\sqrt[3]{\left(S_{d}+S_{a}\right)^{2}S_{d}^{4}S_{a}^{2}}\left[J_{\parallel}^{2}\left\vert J_{\perp}\right\vert \right]^{\frac{1}{3}} \label{Neel3D}
\end{equation}
-- the 3D structure-specific relation of the effective exchange interaction to the 
N{\'e}el temperature.  

The assumptions used in Appendix \ref{SpinWave} for deriving the formula eq. (\ref{Neel3D})
are not satisfied in strongly anisotropic systems where $\left\vert J_{\perp}\right\vert \ll J_{\parallel}$.
In this limit the N{\'e}el temperature is to be determind from the transcendental
equation:
\begin{equation}
\label{Neel2D}
\frac{1}{4\pi}\frac{S_{d}-S_{a}}{S_{d}S_{a}}\frac{\theta_{\mathrm{N}}}{J_{\Vert}}\frac{1}{S_{d}+S_{a}}\log\left(\frac{S_{d}-S_{a}}{S_{d}^{2}}\frac{\theta_{\mathrm{N}}}{\left\vert J_{\bot}\right\vert }\right)=1,
\end{equation}
and the magnetization is given by: 
\begin{equation}
M_{s}=M_{0}\left[1-\frac{1}{4\pi}\frac{S_{d}-S_{a}}{S_{d}S_{a}}\frac{1}{S_{d}+S_{a}}\frac{\theta}{J_{\Vert}}\log\left(\frac{S_{d}-S_{a}}{S_{d}^{2}}\frac{\theta}{\left\vert J_{\bot}\right\vert }\right)\right]
\end{equation}
provided $\theta$ is close enough to $\theta_{\mathrm{N}}$.

Neither three-dimensional (3D) or two-dimensional (2D) estimates of 
the N\'{e}el temperature eqs. (\ref{Neel3D}) and (\ref{Neel2D}), respectively, 
permits to determine the longitudinal and transversal interactions independently and to establish by this the 
amount of anisotropy. 
The effective exchange interaction $J^{\rm SW}_{\rm eff} = \sqrt[3]{J_{\parallel }^2J_{\perp}}$ 
as derived from eq. (\ref{Neel3D}) and the experimental  N\'{e}el temperature 
for the HM V(TCNE)$_2$ compound 
amounts  $J^{\rm SW}_{\rm eff}({\rm V}) = $ 36 K. This is due rather large numerical value of the 
transition coefficient 
in eq. (\ref{Neel3D})
coupling the 
effective exchange interaction with the N\'{e}el temperature (11.376 for the 
structure depicted in Fig. 1 and $S_d = \frac32; S_a = \frac12 $). This result is in a general agreement
with the result of Ref. \cite{WeiQiuDu} which yields the corresponding coefficient 
to be 9.937 for a simple cubic ferrimagnet with the same values of the effective spins. 
(It is not clear how the estimate of {\it ca.} 100 K for  $J^{\rm SW}_{\rm eff}({\rm V})$ 
is obtained in Ref. \cite{PPEM2001} since is also based on assumption 
of a simple cubic lattice magnetic structure, 
but apparently uses some different coefficient). 
It also stresses the different character of averaging of intralayer and interlayer
exchange parameters in the mean field (arithmetic mean) and in the spin-wave (geometric mean)
approximations. At the high anisotropies the geometric mean provides much stronger dependence of the 
effective exchange on the interlayer exchange than the arithmetic mean. 
 
Whatever value of $J^{\rm SW}_{\rm eff}$ leaves a wide range of 
possibilities since each pair of
values of $J_{\perp} $  and $J_{\parallel } $ yielding the above value of $J^{\rm SW}_{\rm eff}({\rm V})$
conforms with the experimental data on magnetization. 
It has to be realized, however, that using the Bloch $T^{\frac{3}{2}}$ law 
for the magnetization 
in the entire temperature range below $T_{N}$ is an extrapolation of the data 
obtained at low temperatures. 
In order to check its validity in a wider temperature range we notice that 
according to it the 
magnetization depletion at the an intermediate temperature (225 K) amounts 
the factor of 0.592 which looks out to be in an acceptable agreement with experiment 
which shows the magnetization depletion 
by a factor of {\it ca.} 0.6 
at this temperature as compared to that at $T=0$. 

In the HM V(TCNE)$_{2}$ case the measured magnetization values 
are available up to 300 K. The Bloch $T^{\frac{3}{2}}$-law for the
temperature dependence of spontaneous magnetization results in a simple formula 
for the slope of the magnetization {\it vs.} temperature in the N\'{e}el point:
\begin{equation}
-\frac32 \frac {M_0}{T_{\rm N}}
\end{equation}%
As one can derive from the magnetization data on the HM compound given in Ref. 
\cite{MillerEpstein} the slope of the magnetization of the HM material in the N\'{e}el point
amounts $\displaystyle -1.4 \frac {M_0}{T_{\rm N}}$ which is a fair extrapolation 
of the last measured points. It suggests the 3D
regime for the HM material in the entire temperature interval up to  the N\'{e}el point. 
Nevertheless the possibility 
of transition to the 2D regime at $T > 300$ K cannot be {\it a priori} excluded. 
Then for the 2D regime the slope 
of magnetization {\it vs.} temperature 
in the N\'{e}el point
 is:
\begin{equation}
-M_0\frac{ k_B}{4\pi J_{\parallel }}\frac{S_d - S_a}{S_d S_a} \frac{1}{S_d + S_a}
\left(1 + \log\left(\frac{S_d - S_a}{S _d ^2}\frac{k_B T_{\rm N}}{ \left\vert J_{\perp } \right\vert }\right)\right)
\end{equation}%
When combined with eq. (\ref{Neel2D}) it yields:
\begin{equation}
-M_0\left(\frac{k_B}{4\pi J_{\parallel }}\frac{S_d - S_a}{S_d S_a} \frac{1}{S_d + S_a} + \frac {1}{T_{\rm N}}\right)
\end{equation}%
Employing the value of the slope extracted from Fig. 1  
of Ref. \cite{MillerEpstein} we derive $k_B T_{\rm N} = 2.4 \pi J_{\parallel}$, and inserting 
experimental (extrapolated) value of $ T_{\rm N}$ yields immediately 
$J_{\parallel} = 54 $ K and together with the value of $J^{\rm SW}_{\rm eff}$  extracted from
low temperature data allows to estimate anisotropy to be
($J_{\parallel }/J_{\perp} \approx 3 $). 
At the above
intermediate temperature (225 K) the 2D estimate with this anisotropy 
shows the depletion of magnetization
 to be 0.595 of the 
maximal value at 0 K as well in a perfect agreement with experiment, which shows that
the available data on the temperature dependence of magnetization in HM V(TCNE)$_2$ do not allow
to distinguish between the 3D and 2D regimes.   
It must be admitted that in general the above value of anisotropy  is not large enough 
($ \sim 3 $) for the 2D regime to install. This analysis, however, allows us to set bounds for the 
value $J_{\parallel }$ in the V(TCNE)$_2$ compound as derived from the spin-wave treatment: it appears that
this compound resides in the 3D regime so that the above value of anisotropy must be considered as a maximal possible
in this material
. Otherwise even higher N\'{e}el temperatures (although not accessible expreimentally due to material's
decomposition) still conforming to the applicability conditions ($2J_{\perp}S_d^2 \ll \theta \ll 4 J_{\parallel }S_d S_a $) Ref. \cite{Katanin-Irkhin} of the logarithmic formula eq. (\ref{Neel2D}) should have to be admitted. 

Applying analogous treatment to the Fe(TCNE)$_2$ compound for which the structure presented
in Fig. 1 is {\em experimentally} established yields the following: $J^{\rm SW}_{\rm eff} = 9.5$ K 
(the coefficient of 12.914 coming from eq. (\ref{Neel3D}) with $S_d = 2$ is used). With the anisotropy
of $2.5^3 = 15.625$ the 2D estimate of the N\'{e}el temperature is 124 K again in a fair agreement with
the experiment. The value of $J_{\parallel }$ is then 24 K. This turns out to be
not that much different from the upper boundary for the same quantity for the 
V(TCNE)$_2$ compound yielding the ratio of the intralayer exchange parameters for the 
two materials of maximum only two, instead of 4 $\div $ 5 stipulated
by the mean field estimates, and only 1.5 if the isotropic regime is accepted for 
the V(TCNE)$_2$ compound. 

This latter value can be fairly explained by addressing the formulae
of Appendix \ref{SpinHamiltonian} and the spectroscopic data. From eq. (\ref{Heisenberg-exchange-parameter})  
it follows that the ratio $J_{\parallel }({\rm V})/J_{\parallel }({\rm Fe})$ 
 of the intralayer
parameters for the vanadium and iron compounds
is that of the squared hopping parameters $t_{da}$. (We assume here that due to similarity 
of the environment in these two compounds the energy denominators in eq. (\ref{Heisenberg-exchange-parameter})  
given by eq. (\ref{EnergyDenominators}) 
are the same for the both compounds since the values of 
ionisation potentials 
of the V$^{2+}$ and Fe$^{2+}$ ions which are respectively 29.55 and 30.90 eV 
as coming from Ref. \cite{NIST} and similarly close estimates for the electron affinities for these ions). 
According to suggestion by \cite{Anderson} thoroughly tested numerically in Refs. \cite{Tch017,Tch033,Tch037} the amounts of the crystal field splitting in the 
coordination compounds are proportional to analogous expressions: squares of the hopping parameters
divided by some (other) energy denominators, which are, however, also approximately equal in
similar compounds. Thus the proportion holds: 
\[
\frac{J_{\parallel}({\rm V})}{J_{\parallel}({\rm Fe})}  = \frac{10Dq({\rm V})}{10Dq({\rm Fe})}
\]
for pairs of similar complexes in each side of the proportion. For the right side of the proportion we find with the
values of Refs.  
\cite{Hitchcock,Lever} $10Dq({\rm V}) = 14700 $ cm$^{-1}$, $10Dq({\rm Fe}) = 10900 $ cm$^{-1}$
to be 
1.35 for the hexacoordinate octahedral complexes with acetonitile, which 
basically explains the above ratio 1.5 of the intralayer exchange parameters. Of course,
the significant difference of anisotropies in these two materials remains to be understood. 

This can be 
tentatively done with use of the Goodenough-Kanamori rules \cite{Goodenough}. Indeed, the difference in 
the interlayer interactions requiring an explanation is too large, so that probably a qualitative 
distinction between the two materials is responsible for it. As we mentioned above the simplistic 
application of the Goodenough-Kanamori rules in the present situation yields an {\em antiferromagnetic}
sign of the interlayer interaction (the situation falls into the Goodenough-Kanamori cation-anion-cation category in the 
180$^{\circ}$ geometry). Thus the observed {\em ferromagnetic} sign of the interlayer interaction appears as
a result of ferromagnetic contributions of the higher order. Such contributions depend qualitatively on the 
possibility to take advantage of the intrashell ferromagnetic interactions which in their turn depend on 
the occupancies of the atomic orbitals in the d-shells of the interacting transition metal cations. 
(Importance of such terms in the context of organometallic magnets had been stressed in Refs. 
\cite{Tch018,Tch019}). It can be easily understood that the conditions for appearance of the 
compensating ferromagnetic terms are very much different for the V$^{2+}$ and Fe$^{2+}$ ions. Indeed, in the case 
of the V$^{2+}$ ion 
two $d$-orbitals remain empty and can participate in the one-electron transfer coupled with
the intrashell exchange eq.(\ref{dd-ferro}) compensating otherwise dominating antiferromagnetic
kinetic exchange. In the case of the Fe$^{2+}$ ion only one  {\em doubly} occupied $d$-orbital can take part in a 
similar process, so that one can expect that the compensating contribution will be significantly weaker
in the case of the Fe(TCNE)$_2$ compound eventually leading to much weaker overall {\em ferromagnetic}
interlayer interaction, than in the case of the V(TCNE)$_2$ compound. 

\section{Discussion}

In the present Section we apply the models proposed above to analysis of 
experimantal data available for the HM V(TCNE)$_{2}$ and Fe(TCNE)$_{2}$ compounds. 
 
First of all we notice that the spin polarization per unit cell (number of
electrons with spin up minus that with spin down) which can be related with
observed magnetization per formula unit. We see that the
calculation performed for V(TCNE)$_{2}$ at the experimental structure 
of Fe(TCNE)$_{2}$ depicted on 
Fig. 1 
shows the spin
polarization of {\it ca.} 8 spins-1/2 per unit cell corresponding to two netto
unpaired electrons per formula unit which is in a fair agreement with the magnetization
measured in the HM V(TCNE)$_{2}$ compound. On the other hand
the LM V(TCNE)$_{2}$ material manifests a weaker
saturation magnetization, namely corresponding to {\it ca. } one netto unpaired
electron per formula unit. This allows to think about certain differences in the
structures of two materials. Nevertheless, both experimentally observed values ({\it ca.} 10$%
\cdot $10$^{3}$ emu$\cdot $Oe$\cdot $mol$^{-1}$ and 6$\cdot $10$^{3}$ emu$%
\cdot $Oe$\cdot $mol$^{-1}$, respectively) both deviate from the theoretical
values of 11.2 and 5.6 giving the magnetization produced by the integer
number of netto spin-polarized electrons in an assumption of the Land\'{e} factor
being equal to 2. 


When trying to extend the model Ref. \cite{Her} of the Fe(TCNE)$_2$ compound to 
analysis of the V(TCNE)$_2$ 
compound the authors Ref. \cite{Her} argued that the vanadium compound must have some
structure different from the iron one since the saturation magnetization in it is lower and
approximately corresponds to two spins $1/2$ compensating (interacting
antiferromagnetically with) one spin $3/2$ per formula unit. From this
observation the authors of Ref. \cite{Her} conclude that the interlayer
interactions must be mediated by $\mu _{4}$-TCNE radical-anions, as it has
been suggested yet in \cite{Tchougreeff-Hoffmann}, rather by the 
[TCNE]$_2^{2-}$ dimers. 
This argument applies of course only to the LM form of the V(TCNE)$_2$ material
since for its HM form our numerical experiment shows that for the
experimental structure of the Fe(TCNE)$_2$ compound the calculated magnetization fairly
corresponds to the experimental value obtained on the HM V(TCNE)$_2$ material. 
Incidentally, the magnetization values obtained numerically at
intermediate structures on the \ "reaction path" depicted on Fig. 3 
allows us to assume that some similar structures obtained from 
structures of Fig. 2 
by rotations of some TCNE units may present in
the LM V-TCNE material. This view had found a 
recent support from the computational side in Ref. \cite{DeFusco} where a structure 
for the LM form of V(TCNE)$_2$ has been proposed. It would be fair to say (although
it is not said in Ref. \cite{DeFusco}) that this structure as well descends from the 
structures of Refs. \cite{Tchougreeff-Hoffmann, Tch082}. Specifically,
in order to obtain the structure of Ref. \cite{DeFusco} one has to  rotate
 the TCNE molecule laying in the $bc$-face 
of either of these structures depicted in Fig. 2 
in each of the unit cells 
around the diagonal of the face (or around 
an axis going through the pair of {\it trans}-nitrogen atoms of that TCNE unit) 
by {\it ca. } 90$^\circ$ so that two other N-atoms of each rotating TCNE unit go out of
coordination with the V ions. Such a structure corresponds as that of Ref. \cite{Tchougreeff-Hoffmann}
and the principal one (see above) of Ref. \cite{Tch082} to two TCNE$^{\dot{-}}$ units per 
unit cell each bearing one unpaired electron and thus expectedly yield the overall 
magnetization corresponding to one unpaired electron per formula unit. For such a structure
the magnetic interaction parameters obtained in Ref. \cite{DeFusco} are almost isotropic 
($J_{\perp} = 720 $ K and $J_{\parallel } = 690  $ K) which is also not surprising 
since the character of interactions between the V $d$-shells and LUMO's of the TCNE$^{\dot{-}}$ 
units are fairly the same in either direction. The numerical values of the exchange 
parameters obtained in Ref. \cite{DeFusco} are for sure 
considerable overestimates of the true ones since the N{\'e}el temperature derived from them either
by the mean field or spin-wave methods exceeds the experimental value by orders of magnitude.
Nevertheless, applying the spin-wave theory similar to that described in Section \ref{SpinWave}
results in the estimate for $k_B T_{\rm N} = 9.14 J^{\rm SW}_{\rm eff}$ which yields the numerical value of 
 $J^{\rm SW}_{\rm eff}$ for the LM form of V(TCNE)$_2$ material of 45 K in fair agreement with the similar above estimates for the
intralayer effective exchange parameters.  
It must also
be admitted that the spin-wave treatment of the model of Ref. \cite{DeFusco} leaves the question of the 
reason of complete disagreement of the temperature dependence of the magnetization in the 
LM compound as given in Ref. \cite{MillerEpstein} with the Bloch law which should be expected for almost isotropic 
ferrimagnet unanswered.  

\section{Conclusion}

In the present paper we performed detailed analysis of our numerical results
concerning thinkable structure of room-temperature organometallic magnet
V(TCNE)$_{2}$ as manifested in the corresponding projections of DoS. Similar
analysis of projected DoS for a sequence of structures leading to the
tentative experimental structure of V(TCNE)$_{2}$ is performed as well.
Model spin Hamiltonian is developed for analysis and
interpretation of numerical results and experimental data. Analysis of
magnetic data in terms of the approximate models derived from the 
phenomenological Hamiltonian is performed. A remarkable
correspondence between experimental (structural and magnetic) data on V(TCNE)%
$_{x}\cdot y\ {\rm solvent}$ and numerical model has been observed previously:
magnetization corresponding to two unpaired electrons per formula unit in fair
agreement with experiment on HM V-TCNE material derived from V(CO)$_{6}$ by CVD
technique is obtained numerically for V(TCNE)$_{2}$ taken in the relaxed
experimental Fe(TCNE)$_{2}$ geometry Ref. \cite{Tch082}. Now it is complemented by 
the detailed analysis of the magnon spectrum of this model. The possible transition 
between the low-temperature 3D and the high-temperature 2D regimes is discussed. 
Estimates
of parameters of the proposed spin-Hamiltonian as treated in the spin-wave
approximation are derived from the experimental data on
the N\'{e}el temperature and the temperature dependence of magnetization. The differences in 
magnetic behavior of probably isostructural HM V(TCNE)$_{2}$ and Fe(TCNE)$_{2}$ are
tentatively explained. 

\section*{Acknowledgments}

This work is performed with the partial support of the RFBR grant No
07-03-01128 extended to ALT. The generous support of the visit and stay of \
ALT at RWTH -- Aachen University by DFG through the grant No DR 342/20-1 is
gratefully acknowledged. ALT is thankful to Dr B. Eck for his help in
mastering the solid state electronic structure analysis software at the IAC
of the RWTH and to Drs. A.M. Tokmachev, I.V. Pletnev, and 
J. von Appen for valuable discussions. Prof. Joel S.
Miller is acknowledged for kindly drawing the authors' attention to Ref. \cite{DeFusco} and for
sending a preprint of his work \cite{MillerPreprint} prior to publication. 
\appendix
\section{Spin Hamiltonian as derived from orbital model Hamiltonian}\label{SpinHamiltonian}

Simple estimates of the parameters of the spin Hamiltonian eq, (\ref{true-spin-Hamiltonian})
can be based on considerations dating back to Ref. \cite{Goodenough} as specified for the 
current situation. For the idealized geometry of  M(TCNE)$_{2}$ compounds where the  M-TCNE layers
are assumed to be planar the one-electron hopping parameters $t_{da}$ are those between the 
$b_{3g}\left( \pi ^{\ast }\right) $ singly occupied orbital of the TCNE$^{\dot{-}}$ radial-anion 
and one of the $d$-orbitals of the $\pi$-symmetry with respect to the abovementioned plane
($d_{xz}$ and $d_{yz}$). 
For a pair of the  TCNE$^{\dot{-}}$ and V$^{2+}$ ions described by
the Hamiltonians, eqs. (\ref{a-Hamiltonian}) and (\ref{d-Hamiltonian}),
respectively, the energy of the bare ground state reads: 
\begin{equation}
E_{0}=-\alpha _{a}-3\alpha _{d}+3U_{dd}-21/4 J_{dd}
\label{pair-ground-state-energy}
\end{equation}%
and does not depend on the way the spins in these sites are coupled. The one-electron hopping
couples them with
two states with one electron transferred between the two sites (from one of
the $d$-states to the $a$-state and from the $a$-state to one of the $d$%
-states). The energies of the
charge transfer states
are respectively: 
\begin{eqnarray*}
E_{d\rightarrow a} &=&-2\alpha _{a}-2\alpha _{d}+U_{aa}+U_{dd}-3J_{dd}/2 \\
E_{a\rightarrow d} &=&-4\alpha _{d}+6U_{dd}+29/4J_{dd}
\end{eqnarray*}%
They both correspond to the lower spin of the state of the two ions which
results, as usual, to a Heisenberg-type interaction of two 1/2 electron
spins: 
\begin{equation}
2K_{da}\hat{S}_{\gamma {\bf r}}\hat{S}_{a{\bf r}}
\label{Heisenberg-exchange}
\end{equation}%
occupying the overlapping orbitals (here $\gamma $ refers to that of the
three $d$-orbitals which overlaps with the particular $a$-orbital) with the
effective exchange constant given by: 
\begin{equation}
K_{da}=t_{da}^{2}(\Delta E_{d\rightarrow a}^{-1}+\Delta E_{a\rightarrow
d}^{-1})>0,  \label{Heisenberg-exchange-parameter}
\end{equation}%
where 
\begin{eqnarray*}
\Delta E_{d\rightarrow a} &=&\alpha _{d}-\alpha
_{a}+U_{aa}-2U_{dd}+15/4 J_{dd}=\Delta \alpha +U_{aa}-2U_{dd}+15/4 J_{dd}>0, \\
\Delta E_{a\rightarrow d} &=&\alpha _{a}-\alpha _{d}+3U_{dd}+50/4J_{dd}=-\Delta
\alpha +3U_{dd}+50/4 J_{dd}>0, \\
\Delta \alpha  &=&\alpha _{d}-\alpha _{a}>0;\Delta E_{a\rightarrow d}>\Delta
E_{d\rightarrow a}.
\end{eqnarray*}%
On the other hand the charge transfer energies can be expressed through
 the spectral ionization
potentials of the respective ions, their electron affinities and the energy shifts
$C_d$ and $C_a$
of these quantities induced by the  Coulomb field of the surrounding crystals already
mentioned: 
\begin{eqnarray}
\label{EnergyDenominators}
\Delta E_{d\rightarrow a} &=&I _{d}-A_{a}-g_{ad}>0, ,\nonumber \\
\Delta E_{a\rightarrow d} &=&I _{a}-A _{d}-g_{ad}>0, \nonumber \\
I _{d}  &=&I^0 _{d}-C _{d}>0; I^0 _{d} = \alpha ^0 _{d} - (n_d - 1)U_{dd} \\
A^0 _d &=& I ^0 _{d} - U_{dd} \nonumber \\
I _{a} &=& \alpha ^0 _{a} - C _{a}>0 
. \nonumber
\end{eqnarray}%
(here we omitted the intraatomic exchange parameters $J_{dd}$ known to be by orders of
magnitude smaller than other quantities relevant here and included the electron-hole
interaction energies $g_{ad}$ previously absorbed in $\alpha$'s). 

The interaction eq. (\ref{Heisenberg-exchange}) must be repeated for each
interacting pair of electronic spins (pair of orbitals, coupled by the
 electron hopping operator). For the V ion in the crystal the terms
appear for the $\xi$- and $\eta $-states on each atom. 

For the pair of metal ions interacting through the [TCNE]$_{2}^{2-}$ unit the 
one electron hopping is effectively possible not only between the states in the 
$t_{2g}$-manifolds of the ions involved, but also between other remaining $d$-orbitals. 
By this the states admixed to the spin degenerate bare gound state of the pair of
ions may have either lower or higher spin which leads both to the antiferromagnetic 
contribution of the form:
\begin{equation}
\label{dd-antiferro}
\frac{4t_{dd}^{2}}{\Delta E_{d\rightarrow d}}
\end{equation}
and of the ferromagnetic contribution of the form:
\begin{equation}
\label{dd-ferro}
-\left(\frac{4t_{dd}^{\prime}}{\Delta E_{d\rightarrow d}}\right)^2 J_{dd}
\end{equation}
where the hopping parameters $t_{dd}$ and $t_{dd}^{\prime}$ have the same order of
magnitude. The terms of the antiferromagnetic sign appear for each pair of the  
$d$-orbitals coupled by the hopping, whereas the terms of the ferromegnetic sign 
appear all singly occupied orbitals in both shells provided an empty or a doubly 
occupied in the given $d$-shell couples with a singly occupied orbital in other 
 $d$-shell through the hopping. 
\section{Phenomenological Hamiltonian for effective spins and its relation
to the spin Hamiltonian}\label{SpinToPhen}

The phenomenological Hamiltonian written in terms of the effective spins eq.
(\ref{effective-spins}) to be used for modeling the entire crystal has the form of 
eq. (\ref{phen-Hamiltonian})

In order to obtain it 
we follow the general
recipes given in Ref. \cite{Vonsovsky} for obtaining the spin-wave spectrum 
and write the general Heisenberg equations of
motion for the spin-raising operators $\hat{S}_{d{\bf r}}^{+},\hat{S}_{a{\bf r}}^{+}$. They 
read: 
\begin{equation}
\begin{array}{lll}
i\hbar \frac{\partial }{\partial t}\hat{S}_{d{\bf r}}^{+} & = & -J_{\bot
}\left( \hat{S}_{d{\bf r}}^{z}\hat{S}_{d{\bf r}\pm {\bf b}}^{+}-\hat{S}_{d%
{\bf r}}^{+}\hat{S}_{d{\bf r}\pm {\bf b}}^{z}\right)  \\ 
&  & +J_{\Vert }\left[ \hat{S}_{d{\bf r}}^{z}\left( \hat{S}_{a{\bf r}}^{+}+%
\hat{S}_{a{\bf r}+{\bf a}+{\bf c}}^{+}+\hat{S}_{a{\bf r}+{\bf a}}^{+}+\hat{S}%
_{a{\bf r}+{\bf c}}^{+}\right) \right.  \\ 
&  & -\left. \hat{S}_{d{\bf r}}^{+}\left( \hat{S}_{a{\bf r}}^{z}+\hat{S}_{a%
{\bf r}+{\bf a}+{\bf c}}^{z}+\hat{S}_{a{\bf r}+{\bf a}}^{z}+\hat{S}_{a{\bf r}%
+{\bf c}}^{z}\right) \right]  \\ 
i\hbar \frac{\partial }{\partial t}\hat{S}_{a{\bf r}}^{+} & = & J_{\Vert }
\left[ \hat{S}_{a{\bf r}}^{z}\left( \hat{S}_{d{\bf r}}^{+}+\hat{S}_{d{\bf r}-%
{\bf a}-{\bf c}}^{+}+\hat{S}_{d{\bf r}-{\bf a}}^{+}+\hat{S}_{d{\bf r}-{\bf c}%
}^{+}\right) \right.  \\ 
&  & -\left. \hat{S}_{a{\bf r}}^{+}\left( \hat{S}_{d{\bf r}}^{z}+\hat{S}_{d%
{\bf r}-{\bf a}-{\bf c}}^{z}+\hat{S}_{d{\bf r}-{\bf a}}^{z}+\hat{S}_{d{\bf r}%
-{\bf c}}^{z}\right) \right] 
\end{array}%
\end{equation}%
In the spin-wave approximation the operators $\hat{S}_{d}^{z}$ and $\hat{S}%
_{a}^{z}$ are replaced by their average values in the magnetic (ordered)
phase: 
\begin{equation}
\hat{S}_{d{\bf r}}^{z}\rightarrow \left\langle \hat{S}_{d}^{z}\right\rangle =\frac{3%
}{2};\hat{S}_{a{\bf r}}^{z}\rightarrow \left\langle \hat{S}_{a}^{z}\right\rangle =-%
\frac{1}{2}
\end{equation}%
so that the equations of motion get the form:%
\begin{equation}
\begin{array}{lll}
i\hbar \frac{\partial }{\partial t}\hat{S}_{d{\bf r}}^{+} & = & -J_{\bot
}\left( \frac{3}{2}\hat{S}_{d{\bf r}\pm {\bf b}}^{+}-3\hat{S}_{d{\bf r}%
}^{+}\right)  \\ 
&  & +J_{\Vert }\left[ \frac{3}{2}\left( \hat{S}_{a{\bf r}}^{+}+\hat{S}_{a%
{\bf r}+{\bf a}+{\bf c}}^{+}+\hat{S}_{a{\bf r}+{\bf a}}^{+}+\hat{S}_{a{\bf r}%
+{\bf c}}^{+}\right) +2\hat{S}_{d{\bf r}}^{+}\right]  \\ 
i\hbar \frac{\partial }{\partial t}\hat{S}_{a{\bf r}}^{+} & = & J_{\Vert }
\left[ -\frac{1}{2}\left( \hat{S}_{d{\bf r}}^{+}+\hat{S}_{d{\bf r}-{\bf a}-%
{\bf c}}^{+}+\hat{S}_{d{\bf r}-{\bf a}}^{+}+\hat{S}_{d{\bf r}-{\bf c}%
}^{+}\right) -6\hat{S}_{a{\bf r}}^{+}\right] .%
\end{array}%
\end{equation}%
Going to the Fourier transforms of the raising operators we get:%
\begin{equation}
\label{EOMforSpinRaising} 
\begin{array}{lll}
i\hbar \frac{\partial }{\partial t}\hat{S}_{d{\bf k}}^{+} & = & -3J_{\bot
}\left( \cos k_{b}-1\right) \hat{S}_{d{\bf k}}^{+}+J_{\Vert }\left[
\frac32\Omega _{{\bf k}}\hat{S}_{a{\bf k}}^{+}+2\hat{S}_{d{\bf k}}^{+}\right]  \\ 
i\hbar \frac{\partial }{\partial t}\hat{S}_{a{\bf k}}^{+} & = & -J_{\Vert }
\left[ \frac{\Omega _{{\bf k}}^{\ast }}{2}\hat{S}_{d{\bf k}}^{+}+2\hat{S}_{a{\bf r}%
}^{+}\right] \\
\Omega _{{\bf k}} & = & 1+\exp (i k_{a})+\exp (i k_{c})+\exp (i%
 k_{a}+i k_{c})%
\end{array}
\end{equation}%
which must be complemented by analogous system of equations for the spin lowering
operators $\hat{S}_{d{\bf r}}^{-},\hat{S}_{a{\bf r}}^{-}$.

On the other hand
the Heisenberg equations of motion for the operators $\hat{S}_{\zeta{\mathbf r}}^{+},\hat{S}_{\eta{\mathbf r}}^{+},\hat{S}_{\xi{\mathbf r}}^{+},$
and $\hat{S}_{a{\mathbf r}}^{+}$ 
as derived from eq. (%
\ref{true-spin-Hamiltonian}) 
(after the Fourier transformation is
performed) form a system of four equations of motion for the Fourier
components of the spin-$\frac12$ raising operators for each wave vector ${\mathbf k}$.
It can be rewritten with use of ${\mathbf k}$-dependent 4$\times$4 matrices
acting on the vectors $\hat{S}_{{\mathbf k}}^{+}$ with the components
$\hat{S}_{\zeta{\mathbf k}}^{+},\hat{S}_{\eta{\mathbf k}}^{+},\hat{S}_{\xi{\mathbf k}}^{+},\hat{S}_{a{\mathbf k}}^{+}$:
\begin{equation}
i\hbar\frac{\partial}{\partial t}\hat{S}_{{\mathbf k}}^{+}=M({\mathbf k})\hat{S}_{{\mathbf k}}^{+}\end{equation}
 where \begin{equation}
M({\mathbf k})=A+T({\mathbf k})+L({\mathbf k})\end{equation}
 and the matrix \begin{equation}
A=J_{dd}\left(\begin{array}{cccc}
4 & -2 & -2 & 0\\
-2 & 4 & -2 & 0\\
-2 & -2 & 4 & 0\\
0 & 0 & 0 & 0\end{array}\right)\end{equation}
 describes the spin fluctuations in the $d$-shells, the matrix \begin{equation}
T({\mathbf k})=K_{dd}\left(\cos{\mathbf k}_{b}\left(\begin{array}{cccc}
1 & 1 & 1 & 0\\
1 & 1 & 1 & 0\\
1 & 1 & 1 & 0\\
0 & 0 & 0 & 0\end{array}\right)-3\left(\begin{array}{cccc}
1 & 0 & 0 & 0\\
0 & 1 & 0 & 0\\
0 & 0 & 1 & 0\\
0 & 0 & 0 & 0\end{array}\right)\right)\end{equation}
 describes the spin wave propagation in the $b$-direction (transversal
to the V-TCNE) planes and the matrix

\begin{equation}
L({\mathbf k})=K_{da}\left(\begin{array}{cccc}
0 & 0 & 0 & 0\\
0 & 1 & 0 & \frac{1}{2}Q_{c{\mathbf k}}\\
0 & 0 & 1 & \frac{1}{2}Q_{a{\mathbf k}}\\
0 & -\frac{1}{2}Q_{c{\mathbf k}}^{\ast} & -\frac{1}{2}Q_{a{\mathbf k}}^{\ast} & -2\end{array}\right)\end{equation}
 with 
\begin{equation}
\begin{array}{rcl}
Q_{a{\mathbf k}} & = & 1+e^{-ik_{a}-ik_{c}}\\
Q_{c{\mathbf k}} & = & e^{-ik_{a}}+e^{-ik_{c}} 
\end{array}
\end{equation}
desribes the spin-wave propagation in the $ac$-plane \textit{i.e.}
in the individual V-TCNE layer. Going to the linear combitations of
the spin fluctuation operators in the $d$-shell: $\hat{S}_{\zeta{\mathbf k}}^{+}+\hat{S}_{\eta{\mathbf k}}^{+}+\hat{S}_{\xi{\mathbf k}}^{+};\ -\hat{S}_{\zeta{\mathbf k}}^{+}+\hat{S}_{\eta{\mathbf k}}^{+};\ -\hat{S}_{\zeta{\mathbf k}}^{+}+\hat{S}_{\xi{\mathbf k}}^{+}$
introduces the fluctuation of the effective spin of the $d$-shell
(the first combination) and incidentaly diagonalizes the sum of the
first two matrix terms $A+T({\mathbf k})$ yielding the eigenvalues: $0;\ 3K_{dd}\left(\cos{\mathbf k}_{b}-1\right);\ 6J_{dd}-3K_{dd}>0$
of which the zero one refers to precession of the $\frac{1}{2}$ spin
in the acceptor orbital, the next one refers to precession of the
total spin of the $d$-shell, and the doubly degenerate highest eigenvalue
corresponds to excitations changing the total spin of the $d$-shell.
The interaction matrix $L({\mathbf k})$ in this basis acquires the form:
\begin{eqnarray}
\frac{K_{da}}{6}\left(\begin{array}{llll}
-12 & Q_{c{\mathbf k}}^{*}-2Q_{a{\mathbf k}}^{*} & Q_{a{\mathbf k}}^{*}-2Q_{c{\mathbf k}}^{*} & -Q_{a{\mathbf k}}^{*}-Q_{c{\mathbf k}}^{*}\\
3Q_{a{\mathbf k}} & 4 & -2 & 2\\
3Q_{c{\mathbf k}} & -2 & 4 & 2\\
3\left(Q_{a{\mathbf k}}+Q_{c{\mathbf k}}\right) & 2 & 2 & 4\end{array}\right)\end{eqnarray}
The states corresponding to the excitations of the $d$-shell (the
above matrix is written so that they are the second and the third
ones) are of much higher energy than the states corresponding to precession
of the effective spins in two types of sites of the model. For that
reason the former can be excluded. To do so we treat $L({\mathbf k})$
as a perturbation with the smallness parameter $\frac{K_{da}}{J_{dd}}$
and in the zero order we obtain an operator acting in the two-dimensional
subspace spanned by the vector with the components $\hat{S}_{\zeta{\mathbf k}}^{+}+\hat{S}_{\eta{\mathbf k}}^{+}+\hat{S}_{\xi{\mathbf k}}^{+}=\hat{S}_{d{\mathbf k}}^{+};\ \hat{S}_{a{\mathbf k}}^{+}$:
\begin{equation}
\left(\begin{array}{ll}
-2K_{da} & -\frac{1}{6}\left(Q_{a{\mathbf k}}^{*}+Q_{c{\mathbf k}}^{*}\right)K_{da}\\
\frac{1}{2}\left(Q_{a{\mathbf k}}+Q_{c{\mathbf k}}\right)K_{da} & \frac{2K_{da}}{3}-3K_{dd}\left(1-\cos k_{b})\right)\end{array}\right)
\label{ZeroOrderinJdd}
\end{equation}
Comparing eq. (\ref{ZeroOrderinJdd}) with eq. (\ref{EOMforSpinRaising})
and noticing that $\Omega_{k}=Q_{a{\mathbf k}}^{*}+Q_{c{\mathbf k}}^{*}$
we arrive to the conclusion that they coincide after setting
$\left\vert S_{\xi}^{z}\right\vert =\left\vert S_{\eta}^{z}\right\vert =\frac{1}{2}=\frac{1}{3}\left\vert S_{d}^{z}\right\vert $
and $K_{dd}=-J_{\bot};K_{da}=3J_{\Vert}$ which establishes the required
relation between the spin wave treatments of the Hamiltonians eq.
(\ref{true-spin-Hamiltonian}) and eq. (\ref{phen-Hamiltonian}).
The first order correction to eq. (\ref{ZeroOrderinJdd}) has the form: \begin{eqnarray}
\frac{K_{da}^{2}}{9J_{dd}}\left(\begin{array}{ll}
\frac{1}{2}(2-\cos k_{a}-\cos k_{c}+2\cos k_{c}\cos k_{a}) & \frac{1}{12}\Omega_{k}^{*}\\
-\frac{1}{4}\Omega_{k} & \frac{1}{3}\end{array}\right),\end{eqnarray}
but it is not used hereinafter. 
\section{Mean Field treatment of the phenomenological Hamiltonian.}\label{MeanFieldSection}

In order to obtain a structure dependent mean field estimate for the
N{\'e}el temperature of the V-TCNE material we use the method described in
Ref. \cite{Buschow-de-Boer}. Accordingly the N{\'e}el temperature for a system comprising
two types of magnetic centers with effective spins $S_{d}$ and 
 $S_{a}$ so that each center of the with spin $S_{d}$ has
$Z_{dd}$ neighbours of the same type, $Z_{da}$ neighbours with the spin
$S_{a}$ \emph{etc.} with the interactions between the nearest neighbors
of the specific type given by $J_{dd},J_{da}$ \emph{etc.} satisfies
the equation:\[ 
\left|\begin{array}{cc}
J_{dd}Z_{dd}S_{d}\left(S_{d}+1\right)-3\theta_{N}^{MF} & J_{da}Z_{da}S_{d}\left(S_{d}+1\right)\\
J_{da}Z_{ad}S_{a}\left(S_{a}+1\right) & J_{aa}Z_{aa}S_{a}\left(S_{a}+1\right)-3\theta_{N}^{MF} 
\end{array}\right|=0\]
where we used the fact that $J_{ad}=J_{da}$. For the structure represented
in Fig. 1 and modelled by the Hamiltonian eq. (\ref{phen-Hamiltonian}) we set 
$S_{a}=\frac{1}{2}$. Then $Z_{ad}=Z_{da}=4$; $Z_{dd}=2;Z_{aa}=0$
(acceptors do not have acceptor neighbors in this model). In the notation
of eq. (\ref{phen-Hamiltonian}) $J_{dd}=J_{\perp};J_{da}=J_{\parallel}$, so that we obtain:
\[
\theta_{N}^{MF}=\frac{2J_{\perp}}{3}S_{d}\left(S_{d}+1\right)+\sqrt{\frac{J_{\perp}^{2}}{9}S_{d}^{2}\left(S_{d}+1\right)^{2}+\frac{4}{3}J_{\parallel}^{2}S_{d}\left(S_{d}+1\right)}\]
which in the limit of strong anisotropy yields the following mean
field estimate: 
\begin{equation}
\label{MeanFieldTheta}
\theta_{N}^{MF}=\frac{2J_{\perp}}{3}S_{d}\left(S_{d}+1\right)+J_{\parallel}\sqrt{\frac{4}{3}S_{d}\left(S_{d}+1\right)}
\end{equation}
which flows to the the limiting expression used in Ref. \cite{Her} for the single
Fe-TCNE layer with the variance of the factor of two which results
from the different definition of the Hamiltonian in Refs. \cite{Her,Buschow-de-Boer}. 
\section{Spin-wave model of magnetic properties of V(TCNE)$_{2}$}\label{SpinWave}

Now we address the spin-wave theory of the M(TCNE)$_{2}$ ferrimagnets described by the
phenomenological Hamiltonian eq. (\ref{phen-Hamiltonian}) in order to derive its magnetic properties.  
The temperature dependence of magnetization and the N{\'e}el temperature (that 
at each the magnetization of each sublattice vanishes) are controlled by the spectrum of the lowest
 energy excitations: spin waves  (magnons). Calculation of these properties is 
customary performed with use of the Holstein-Primakoff Ref. \cite{Holstein-Primakoff} representation 
of the magnons. This latter had been many times derived for the ferrimagnets with
relatively simple crystal lattice Refs. \cite{WeiQiuDu,QiuZhang}. These derivations are 
based on the Green's function techniques.
The present case differs from those described 
there by a combination of the alternation of the interaction sign ($J_{\Vert } $ and $J_{\bot}$
have opposite signs) and a relatively complex form of the structure factors which makes the 
Green's function too cumbersome and 
also prevents from using directly the the general formulae derived previously. We perform 
the required derivation for the present structure for certainty using the equation of
motion method. It evolves as follows. 
Each term in the phenomenological Hamiltonian eq. (\ref{phen-Hamiltonian}) has the form:
\begin{equation}
J_{12}\hat{S}_{1}\hat{S}_{2}=J_{12}\left[\hat{S}_{1}^{z}\hat{S}_{2}^{z}+\frac{1}{2}\left(
\hat{S}_{1}^{+}\hat{S}_{2}^{-}+\hat{S}_{1}^{-}\hat{S}_{2}^{+}\right)\right]
\end{equation}
If the pair of the above spins couples ferromagnetically ($J_{12}<0$)
the Holstein-Primakoff operators $b_{i}(b_{i}^{+})$ annihilating
(crearing) an elementary excitation (magnon) at the \emph{i}-th site
are introduced by the relations:
\begin{eqnarray}
\hat{S}_{i}^{+} & = & \sqrt{2S_{i}-\hat{n}_{i}}b_{i} \nonumber \\ 
\hat{S}_{i}^{-} & = & b_{i}^{+}\sqrt{2S_{i}-\hat{n}_{i}}\\
\hat{S}_{i}^{z} & = & S_{i}-\hat{n}_{i};\hat{n}_{i}=b_{i}^{+}b_{i} \nonumber
\end{eqnarray}
The operators $b_{i}(b_{i}^{+})$ so defined obey the boson commutation
relations:
\begin{equation}
\left[b_{i}^{+},b_{i^{\prime}}^{+}\right]=\left[b_{i},b_{i^{\prime}}\right]=0;\left[b_{i}^{+},b_{i^{\prime}}\right]=\delta_{ii^{\prime}}
\end{equation}
After being inserted in the above Hamiltonian term
the square roots are expanded and the terms non higher of the second
order in the boson operators $b_{i}$ are kept (linear spin-wave approximation) so that one gets:
\begin{equation}
J_{12}\hat{S}_{1}\hat{S}_{2}=J_{12}S_{1}S_{2}-J_{12}\left(S_{1}\hat{n}_{2}+\hat{n}_{1}S_{2}\right)+J_{12}\sqrt{S_{1}S_{2}}\left(b_{1}^{+}b_{2}+b_{2}^{+}b_{1}\right)\end{equation}
If the pair of effective spins couples antiferromagnetically ($J_{12}>0$)
the Holstein-Primakoff bosons are obtained by a more complex trick
which, first, requires a transformation of the spins residing on the
$m$-th site in the unit cell by the rotation matrices $\omega_{m}$:
\begin{eqnarray}
\tilde{S}_{i} & = & \hat{S}_{i}\omega_{i};\hat{S}_{i}=\omega_{i}^{\dagger}\tilde{S}_{i}
\end{eqnarray}
In the assumption of the common quantization axis for all spins in
the crystallographic unit cell one can select the rotations as follows:
\begin{equation}
\omega_{1}=\begin{pmatrix}1 & 0 & 0\\
0 & 1 & 0\\
0 & 0 & 1\end{pmatrix};\omega_{2}=\begin{pmatrix}1 & 0 & 0\\
0 & -1 & 0\\
0 & 0 & -1\end{pmatrix}\end{equation}
in the basis of the coordinate spin components $x,y,z$ or \begin{equation}
\omega_{1}=\begin{pmatrix}1 & 0 & 0\\
0 & 1 & 0\\
0 & 0 & 1\end{pmatrix};\omega_{2}=\begin{pmatrix}0 & 1 & 0\\
1 & 0 & 0\\
0 & 0 & -1\end{pmatrix}\end{equation}
in the basis of the tensor spin components $+,-,z$. Then the interaction
term rewrites: \begin{equation}
J_{12}\hat{S}_{1}\hat{S}_{2}=-J_{12}\hat{S}_{1}^{z}\tilde{S}_{2}^{z}+\frac{1}{2}J_{12}\left(\hat{S}_{1}^{+}\tilde{S}_{2}^{+}+\hat{S}_{1}^{-}\tilde{S}_{2}^{-}\right)\end{equation}
The boson operators for the transformed spins are then defined by
the same relations as the nontransformed ones which yields the following
interaction term: 
\begin{equation}
J_{12}\hat{S}_{1}\hat{S}_{2}=-J_{12}S_{1}S_{2}+J_{12}\left(S_{1}\hat{n}_{2}+\hat{n}_{1}S_{2}\right)+J_{12}\sqrt{S_{1}S_{2}}\left(b_{1}^{+}b_{2}^{+}+b_{1}b_{2}\right)\end{equation}
 where the anomalous products $b_{1}^{+}b_{2}^{+}$ and $b_{1}b_{2}$
appear.

Described procedures apply to each pair of the interacting spins
($J_{\Vert}>0;J_{\bot}<0$) in the Hamiltonian eq. (\ref{phen-Hamiltonian})
yielding the following:
\begin{eqnarray}
 H_{\rm SW}& = & \sum_{\mathbf{r}}\left(-4J_{\Vert}S_{a}S_{d}+J_{\bot}S_{d}^{2}+4J_{\Vert}\left(\hat{n}_{a\mathbf{r}}S_{d}+S_{a}\hat{n}_{d\mathbf{r}}\right)-2J_{\bot}S_{d}\hat{n}_{d\mathbf{r}}\right. \nonumber \\ 
 &  & +J_{\Vert}\sqrt{S_{a}S_{d}}\left[D_{\mathbf{r}}\left(A_{\mathbf{r}}+A_{\mathbf{r}+\mathbf{a}}+A_{\mathbf{r}+\mathbf{c}}+A_{\mathbf{r}+\mathbf{a}+\mathbf{c}}\right)\right.\\
 &  & +\left.\left.D_{\mathbf{r}}^{+}\left(A_{\mathbf{r}}^{+}+A_{\mathbf{r}+\mathbf{a}}^{+}+A_{\mathbf{r}+\mathbf{c}}^{+}+A_{\mathbf{r}+\mathbf{a}+\mathbf{c}}^{+}\right)\right]+J_{\bot}S_{d}\left(D_{\mathbf{r}}^{+}D_{\mathbf{r}+\mathbf{b}}+D_{\mathbf{r}+\mathbf{b}}^{+}D_{\mathbf{r}}\right)\right) \nonumber
\end{eqnarray}
where $D_{\mathbf{r}}^{+}(D_{\mathbf{r}})$ and $A_{\mathbf{r}}^{+}(A_{\mathbf{r}})$
are the magnon creation (annihilation) operators at the respective
sites in the crystal and the 'numbers of magnons' operators necessary
to calculate the magnetization at each site are: \begin{equation}
\hat{n}_{a\mathbf{r}}=A_{\mathbf{r}}^{+}A_{\mathbf{r}};\hat{n}_{d\mathbf{r}}=D_{\mathbf{r}}^{+}D_{\mathbf{r}}\end{equation}
Introducing the Fourier transforms of the magnon creation operators
by the relations:\begin{eqnarray}
D_{\mathbf{r}}^{+} & = & \frac{1}{\sqrt{N}}\sum_{\mathbf{k}}\exp(-i\bf{kr}) D_{\mathbf{k}}^{+}; \nonumber \\
A_{\mathbf{r}}^{+} & = & \frac{1}{\sqrt{N}}\sum_{\mathbf{k}}\exp(-i\bf{kr}) A_{\mathbf{k}}^{+}\end{eqnarray}
and by the and the hermitean conjugate ones for the annihilation operators we obtain the spin-wave Hamiltonian in the form:
\begin{eqnarray}
H_{\rm SW}& = & \sum_{\mathbf{k}}\left(4J_{\Vert}\left(S_{d}A_{\mathbf{k}}^{+}A_{\mathbf{k}}+S_{a}D_{\mathbf{k}}^{+}D_{\mathbf{k}}\right)\right. \nonumber \\
 &  & +J_{\Vert}\sqrt{S_{a}S_{d}}\left[\Omega_{\mathbf{k}}^{\ast}D_{\mathbf{k}}A_{-\mathbf{k}}+\Omega_{\mathbf{k}}D_{\mathbf{k}}^{+}A_{-\mathbf{k}}^{+}\right] \\
 &  & +\left.2J_{\bot}S_{d}\left(\cos k_{b}-1\right)D_{\mathbf{k}}^{+}D_{\mathbf{k}}\right) \nonumber
\end{eqnarray}
where the unnecessary constant is omitted and the structure factors
$\Omega_{\mathbf{k}}$ coincide with those defined by eq. (\ref{EOMforSpinRaising}).

The above Hamiltonian produces the Heisenberg equations of motion
\[
i\hbar \dot{b} = \left[H_{\rm SW} , b \right]
\]
 for the annihilation operators 
and which are coupled with the analogous ones for the creation operators 
due to the presence of
the anomalous terms:
\begin{eqnarray}
\label{HPEOM}
i\hbar\dot{A}_{\mathbf{k}} & = & 4J_{\Vert}S_{d}A_{\mathbf{k}}+J_{\Vert}\sqrt{S_{a}S_{d}}\Omega_{-\mathbf{k}}D_{-\mathbf{k}}^{+} \nonumber \\
i\hbar\dot{D}_{\mathbf{k}}^{+} & = & -4J_{\Vert}S_{a}D_{\mathbf{k}}^{+}-J_{\Vert}\sqrt{S_{a}S_{d}}\Omega_{\mathbf{k}}^{\ast}A_{-\mathbf{k}}+2J_{\bot}S_{d}\left(\cos k_{b}-1\right)D_{\mathbf{k}}^{+}\\
i\hbar\dot{D}_{\mathbf{k}} & = & 4J_{\Vert}S_{a}D_{\mathbf{k}}+J_{\Vert}\sqrt{S_{a}S_{d}}\Omega_{\mathbf{k}}A_{-\mathbf{k}}^{+}-2J_{\bot}S_{d}\left(\cos k_{b}-1\right)D_{\mathbf{k}} \nonumber \\
i\hbar\dot{A}_{\mathbf{k}}^{+} & = & -4J_{\Vert}S_{d}A_{\mathbf{k}}^{+}-J_{\Vert}\sqrt{S_{a}S_{d}}\Omega_{-\mathbf{k}}^{\ast}D_{-\mathbf{k}} \nonumber
\end{eqnarray}
One can easily see that in the above system
the first one is coupled only to the second whereas the third one
is coupled only with the fourth. Thus eq. (\ref{HPEOM}) reduces to a pair of 2$\times$2
matrix eigenvalue/eigenvector problems to be solved for the stationary
magnons:
\begin{equation}
\label{HPEOM1}
\begin{pmatrix}4S_{d}-\varepsilon_{\mathbf{k}} & \sqrt{S_{a}S_{d}}\Omega_{-\mathbf{k}}\\
-\sqrt{S_{a}S_{d}}\Omega_{-\mathbf{k}}^{\ast} & -4S_{a}+2S_{d}a_{\mathbf{k}}-\varepsilon_{\mathbf{k}}\end{pmatrix}\begin{pmatrix}u_{\mathbf{k}}\\
-v_{\mathbf{k}}\end{pmatrix}=0
\end{equation}
\begin{equation}
\label{HPEOM2}
\begin{pmatrix}4S_{a}-2S_{d}a_{\mathbf{k}}-\varepsilon_{\mathbf{k}} & \sqrt{S_{a}S_{d}}\Omega_{\mathbf{k}}\\
-\sqrt{S_{a}S_{d}}\Omega_{\mathbf{k}}^{\ast} & -4S_{d}-\varepsilon_{\mathbf{k}}\end{pmatrix}\begin{pmatrix}x_{\mathbf{k}}\\
-y_{\mathbf{k}}\end{pmatrix}=0\end{equation}
where \begin{equation}
a_{\mathbf{k}}=\frac{J_{\bot}}{J_{\Vert}}\left(1-\cos k_{b}\right)\end{equation}
The matrix eigenvalue problems eqs. (\ref{HPEOM1}), (\ref{HPEOM2}) each yield two solutions for $\varepsilon_{\mathbf{k}}$
of which one is negative in either case and must be rejected
(see Ref. \cite{Vonsovsky}). The nonegative solutions of both problems
represent the spectrum of magnon excitations having the form:
\begin{eqnarray}
\varepsilon_{\mathbf{k}}^{\pm} & = & \Gamma_{\mathbf{k}}\pm\left(2(S_{d}-S_{a})+S_{d}a_{\mathbf{k}}\right)\\
\Gamma_{\mathbf{k}} & = & \sqrt{\left(2(S_{d}+S_{a})-S_{d}a_{\mathbf{k}}\right)^{2}-S_{d}S_{a}\left\vert \Omega_{\mathbf{k}}\right\vert ^{2}} \nonumber
\end{eqnarray}
where the "$-$" sign corresponds to the gapless
band with the magnon annihilation operators $G_{\mathbf{k}}$, whereas the
\char`\"{}$+$\char`\"{} sign in the expression for the energy corresponds
to the excitations with a gap and the magnon annihilation operators $F_{\mathbf{k}}$
 given by the relations:
\begin{eqnarray}
F_{\mathbf{k}} & = & u_{\mathbf{k}}A_{\mathbf{k}}-v_{\mathbf{k}}D_{-\mathbf{k}}^{+}\\
G_{\mathbf{k}} & = & x_{\mathbf{k}}D_{\mathbf{k}}-y_{\mathbf{k}}A_{-\mathbf{k}}^{+}\end{eqnarray}

The site populations by the magnons are given by
\begin{equation}
\left\langle \hat{n}_{a\mathbf{r}}\right\rangle =\frac{1}{N}\sum_{\mathbf{k}}\left\langle A_{\mathbf{k}}^{+}A_{\mathbf{k}}\right\rangle ;\left\langle \hat{n}_{d\mathbf{r}}\right\rangle =\frac{1}{N}\sum_{\mathbf{k}}\left\langle D_{\mathbf{k}}^{+}D_{\mathbf{k}}\right\rangle \label{SiteMagnonDensity}\end{equation}
where 
\begin{eqnarray}
\left\langle A_{\mathbf{k}}^{+}A_{\mathbf{k}}\right\rangle  & = & \left\vert x_{\mathbf{k}}\right\vert ^{2}\left\langle F_{\mathbf{k}}^{+}F_{\mathbf{k}}\right\rangle +\left\vert y_{\mathbf{k}}\right\vert ^{2}\left\langle G_{\mathbf{k}}G_{\mathbf{k}}^{+}\right\rangle \nonumber \\
\left\langle D_{\mathbf{k}}^{+}D_{\mathbf{k}}\right\rangle  & = & \left\vert v_{\mathbf{k}}\right\vert ^{2}\left\langle F_{\mathbf{k}}F_{\mathbf{k}}^{+}\right\rangle +\left\vert u_{\mathbf{k}}\right\vert ^{2}\left\langle G_{\mathbf{k}}^{+}G_{\mathbf{k}}\right\rangle \end{eqnarray}
and the $\left\langle ...\right\rangle $ averaging is performed over
the equilibium state of a ferrimagnet, which  can be rewritten in compliance with the commutation
rules for the boson operators: 
\begin{eqnarray}
\left\langle A_{\mathbf{k}}^{+}A_{\mathbf{k}}\right\rangle  & = & \left\vert x_{\mathbf{k}}\right\vert ^{2}\left\langle \hat{n}_{F\mathbf{k}}\right\rangle +\left\vert y_{\mathbf{k}}\right\vert ^{2}\left(1-\left\langle \hat{n}_{G\mathbf{k}}\right\rangle \right)
\nonumber \\
\left\langle D_{\mathbf{k}}^{+}D_{\mathbf{k}}\right\rangle  & = & \left\vert v_{\mathbf{k}}\right\vert ^{2}\left(1-\left\langle \hat{n}_{F\mathbf{k}}\right\rangle \right)+\left\vert u_{\mathbf{k}}\right\vert ^{2}\left\langle \hat{n}_{G\mathbf{k}}\right\rangle \end{eqnarray}
The average population of a magnon state at a temperature $T=\theta/k_{B}$
expresses through the energy of the corresponding magnon \begin{equation}
\hbar\omega_{\mathbf{k}}^{\pm}=J_{\Vert}\varepsilon_{\mathbf{k}}^{\pm}\end{equation}
following the boson statistics:\begin{equation}
\left\langle \hat{n}_{G\mathbf{k}}\right\rangle =\left[\exp\left(\frac{\hbar\omega_{\mathbf{k}}^{-}}{\theta}\right)-1\right]^{-1};\left\langle \hat{n}_{F\mathbf{k}}\right\rangle =\left[\exp\left(\frac{\hbar\omega_{\mathbf{k}}^{+}}{\theta}\right)-1\right]^{-1}.\end{equation}
In the low-temperature limit the contribution of the
gap magnons is exponentially small so that $\left\langle \hat{n}_{F\mathbf{k}}\right\rangle $
can be set equal to zero. The expansion amplitudes
$u_{\mathbf{k}},v_{\mathbf{k}},x_{\mathbf{k}},$ and $y_{\mathbf{k}}$
are themselves immaterial and the only required densities  are $\left\vert x_{\mathbf{k}}\right\vert ^{2}$\ and
$\left\vert y_{\mathbf{k}}\right\vert ^{2}$ given by:
\begin{equation}
\left. {{\left\vert x_{\mathbf{k}}\right\vert ^{2}}\atop{\left\vert y_{\mathbf{k}}\right\vert ^{2}}}\right\}
=\frac{1}{S_{d}+S_{a}}\left\{ {S_{d}\atop S_{a}}\right.
\label{Magnon-k-densities}\end{equation}
(here we neglected the $\mathbf{k}$-dependence of the magnon amplitudes). 

The
summation prescribed by eq. (\ref{SiteMagnonDensity}) replaces according
to:\begin{equation}
\frac{1}{N}\sum_{\mathbf{k}}\rightarrow\frac{1}{8\pi^{3}}\int\limits _{BZ}d^{3}\mathbf{k}\end{equation}
with the first Brillouin zone being a cube with the side of $2\pi$.

Evaluating the integrals is based on the advantage of the long wave
approximation for the energy. Two versions of the latter are relevant:
in the first low-temperature regime $\theta\ll\left\vert J_{\bot}\right\vert <J_{\Vert}$
the gapless branch of the magnon energy spectrum has the form: \begin{equation}
\hbar\omega_{\mathbf{k}}^{-}=\left(\frac{S_{d}S_{a}}{S_{d}-S_{a}}J_{\Vert}{\mathbf{k}}_{a}^{2}-\frac{S_{d}^{2}}{S_{d}-S_{a}}J_{\bot}{\mathbf{k}}_{b}^{2}+\frac{S_{d}S_{a}}{S_{d}-S_{a}}J_{\Vert}{\mathbf{k}}_{c}^{2}\right)\end{equation}
which by the standard moves (see \textit{e.g.} Ref \cite{Auerbach})
brings us to the estimates \begin{equation}
\frac{1}{N}\sum_{\mathbf{k}}\left\langle \hat{n}_{G\mathbf{k}}\right\rangle \left\{ {\left\vert x_{\mathbf{k}}\right\vert ^{2}\atop \left\vert y_{\mathbf{k}}\right\vert ^{2}}\right\} =\frac{1}{S_{d}+S_{a}}\left\{ {S_{d}\atop S_{a}}\right\} \frac{\left(S_{d}-S_{a}\right)^{\frac{3}{2}}}{8\pi^{\frac{3}{2}}S_{d}^{2}S_{a}}\zeta\left(\frac{3}{2}\right)\frac{\theta^{\frac{3}{2}}}{J_{\parallel}\left\vert J_{\perp}\right\vert ^{\frac{1}{2}}}\end{equation}
(here $\zeta$ is the Riemann $\zeta$-function). This is in some variance
with Ref. \cite{WeiQiuDu} since in the present case the structure factors
for the magnetic interactions corresponding to the realistic model
of the material under study have been used rather the simple cubic
ones and the definition of the exchange parameters in the Hamiltonian
differs by a factor of two (each pair of interacting spins counts
once here so that the coefficient of the interaction parameter in
the diagonal matrix element of the equation of motion equals to the
number of neighbours of the correcponding type). Taking into account quadratic
corrections to the densities eq. (\ref{Magnon-k-densities}) yields
the higher order corrections $\propto\theta^{\frac{5}{2}}$. Neglecting
these corrections one gets the Bloch $T^{\frac{3}{2}}$ law eq. \ref{BlochMagnetization} for 
the spontaneous magnetization
with the critical (N{\'e}el) temperature (one at which the spontaneous
magnetization disappears and which in the present model coincides
with the point where the sublattice magnetizations disappear either)
to be found from the equiation:\[
\frac{1}{S_{d}+S_{a}}\frac{\left(S_{d}-S_{a}\right)^{\frac{3}{2}}}{8\pi^{\frac{3}{2}}S_{d}^{2}S_{a}}\zeta\left(\frac{3}{2}\right)\frac{\theta_{N}^{\frac{3}{2}}}{J_{\parallel}\left\vert J_{\perp}\right\vert ^{\frac{1}{2}}}=1.
\]

Expanding the magnon energy upto the second order in all components
of the wave vector is possible only if the temperature is the smallest
enery scale. In the case of very high anisotropy the second low temperature
regime is possible when $\left\vert J_{\bot}\right\vert \ll\theta\ll J_{\Vert}$.
In this case the gapless branch of the magnon energy spectrum has
the form: 
\begin{eqnarray}
\hbar\omega_{\mathbf{k}}^{-} & = & \left(\frac{S_{d}S_{a}}{S_{d}-S_{a}}J_{\Vert}k_{p}^{2}-2\frac{S_{d}^{2}}{S_{d}-S_{a}}J_{\bot}\left(1-\cos\mathbf{k}_{b}\right)\right)\\
k_{p}^{2} & = & \mathbf{k}_{a}^{2}+\mathbf{k}_{c}^{2}\nonumber
 \end{eqnarray}
and the integration of $\left\vert x_{\mathbf{k}}\right\vert ^{2}$\ or
$\left\vert y_{\mathbf{k}}\right\vert ^{2}$ divided by $\left[\exp\left(\frac{\hbar\omega_{\mathbf{k}}^{-}}{\theta}\right)-1
\right] $
first performs (following Ref. \cite{SinghTesanovic}) over the planar projection $k_{p}$
of the wave vector $\mathbf{k}$, which reduces to
\begin{equation}
\frac{ab}{4\pi^{2}}\int\limits _{0}^{\infty}\frac{k_{p}dk_{p}}{\left[\exp\left(pk_{p}^{2}\right)-a\right]}\end{equation}
where
\begin{eqnarray}
a & = & \exp\left[\frac{2S_{d}^{2}}{S_{d}-S_{a}}\frac{J_{\bot}}{\theta}\left(1-\cos k_{b}\right)\right];p=\frac{S_{d}S_{a}}{S_{d}-S_{a}}\frac{J_{\Vert}}{\theta};b=\frac{1}{S_{d}+S_{a}}\left\{ {S_{d}\atop S_{a}}\right\} .\end{eqnarray}
After substituting $z=k_{p}^{2}$ the integration yields:
\begin{equation}
-\frac{b}{8\pi^{2}p}\log(1-a)
\end{equation}
Provided $a$ is an exponential function with a small negative argument, 
$\log(1-a)$ replaces by the logarithm of the absolute value of the
argument: 
\begin{equation}
\log(1-a)\approx\log\left(\frac{2S_{d}^{2}}{S_{d}-S_{a}}\frac{\left\vert J_{\bot}\right\vert }{\theta}\left(1-\cos k_{b}\right)\right),
\end{equation}
so that the expression to be intrgrated over $k_{b}$ becomes
\begin{equation}
-\frac{1}{8\pi^{2}}\frac{S_{d}-S_{a}}{S_{d}S_{a}}\frac{\theta}{J_{\Vert}}\frac{1}{S_{d}+S_{a}}\left\{ {S_{d}\atop S_{a}}\right\} \log\left(\frac{2S_{d}^{2}}{S_{d}-S_{a}}\frac{\left\vert J_{\bot}\right\vert }{\theta}\left(1-\cos k_{b}\right)\right)
\end{equation}
which immediately yields\begin{equation}
-\frac{1}{4\pi}\frac{S_{d}-S_{a}}{S_{d}S_{a}}\frac{\theta}{J_{\Vert}}\frac{1}{S_{d}+S_{a}}\left\{ {S_{d}\atop S_{a}}\right\} \log\left(\frac{S_{d}^{2}}{S_{d}-S_{a}}\frac{\left\vert J_{\bot}\right\vert }{\theta}\right).\end{equation}
The sublattice magnetizations are:\[
\left\{ {S_{d}\atop S_{a}}\right\} \left(1-\frac{1}{4\pi}\frac{S_{d}-S_{a}}{S_{d}S_{a}}\frac{1}{S_{d}+S_{a}}\frac{\theta}{J_{\Vert}}\log\left(\frac{S_{d}-S_{a}}{S_{d}^{2}}\frac{\theta}{\left\vert J_{\bot}\right\vert }\right)\right)\]
so that the N{\'e}el temperature is obtained by the condition of the evanescence of the 
sublattice magnetizations which in the present approximation coincide with that of the total 
spontateous magnetization.

\newpage 

\begin{table}
\label{ExpData} 
\center{
\caption{Saturation magnetization $M_s$ at 2 K; ordering temperature $T_c$ and
effective exchange energy $J^{\rm MF}_{\rm eff}$ 
between the metal residing effective spins  
 in the assumption 
of the mean field connection between $T_c$ and $J^{\rm MF}_{\rm eff}$ 
for M(TCNE)$_2$ after \cite{MillerEpstein}.}
\begin{tabular}{|c|c|c|c|c|}
\hline
Metal & Saturation & magnetization $M_s$ & $T_c$ in K & $J_{\rm eff}$ in K \\
      & emu Oe mol-1 & Spins per M atom & & \\
\hline
V$^a$& 10 300&  2& $\sim $ 400& 53 \\
V$^b$& 6 100&  1& $\sim $ 400& - \\
Mn& 19 000&  4& 107& 6.1 \\
Fe& 16 900&  3& 121& 10 \\
Co& 8 000& $\sim $ 1.5& 44& 5.9 \\
Ni& 15 800&  3& 44& 11 \\
\hline
\end{tabular}}\newline
\flushleft{
$^a$ The HM form. \\
$^b$ The LM form.
}
\end{table}

\newpage
\begin{figure}[tbph]
\label{fetcne} 
\center{
\includegraphics*[scale=0.4]{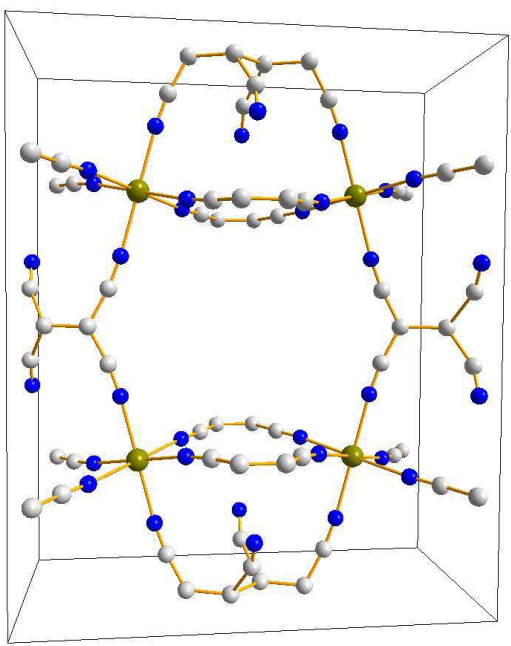}
\caption{Structure of Fe(TCNE)$_{2}$ as coming from the synchrotron radiation study
\protect\cite{Her}. Each unit cell contains four formula units. The solvent
molecules are omitted for clarity.}
}
\end{figure}
\newpage
\begin{landscape}
\begin{figure}[tbp]
\label{str1993} 
\vspace{-2cm}
\includegraphics*[scale=0.7]{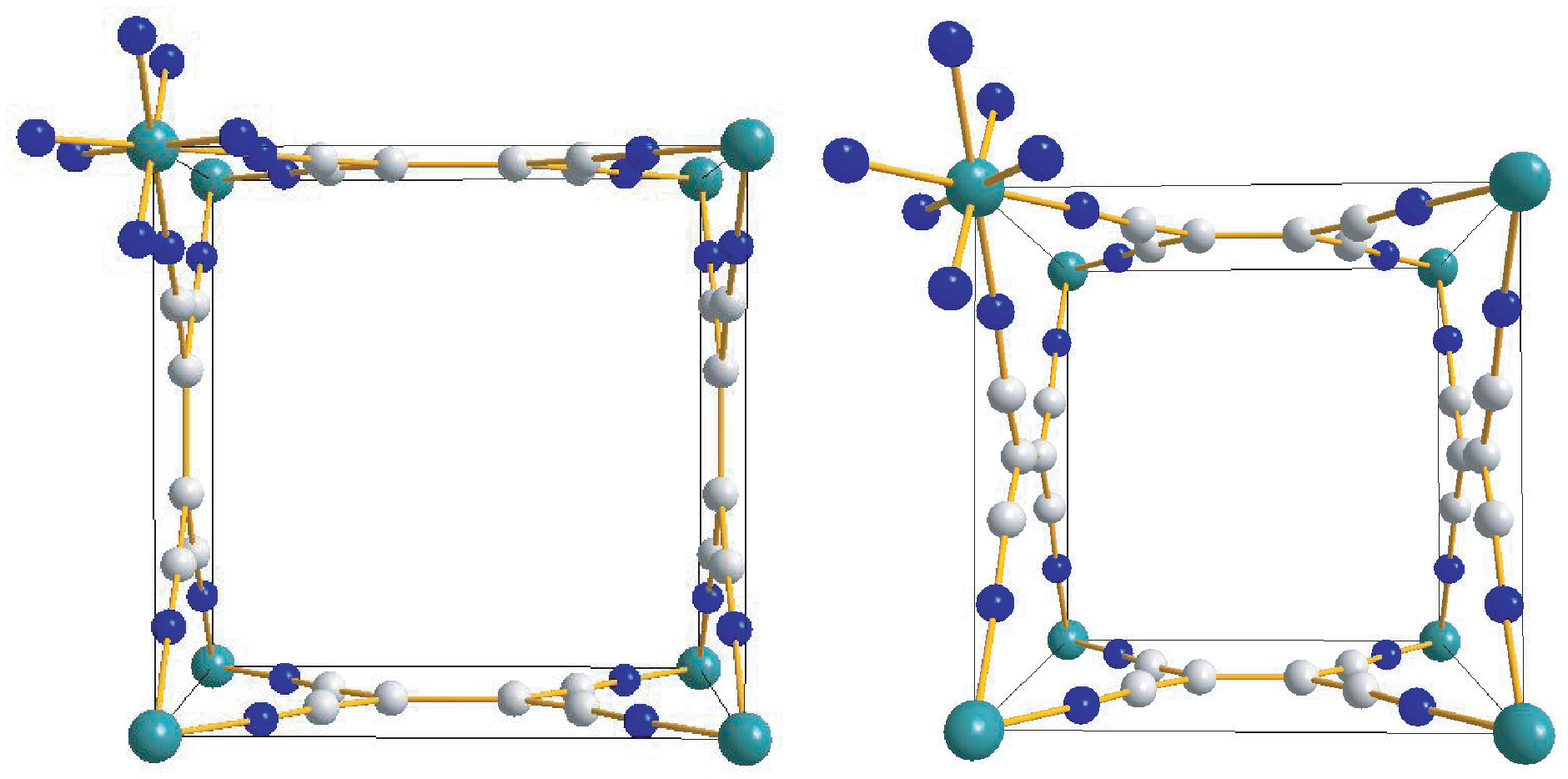}
\caption{Hypothetical structures of V(TCNE)$_{2}$ following \cite{Tchougreeff-Hoffmann} (left) 
and following \cite{Tch082} (right).}
\end{figure}
\end{landscape}
\begin{figure}[tbph]
\label{str0-str10} \vspace{-2cm} 
\includegraphics[bb = 20 20 575 805,scale=0.2]{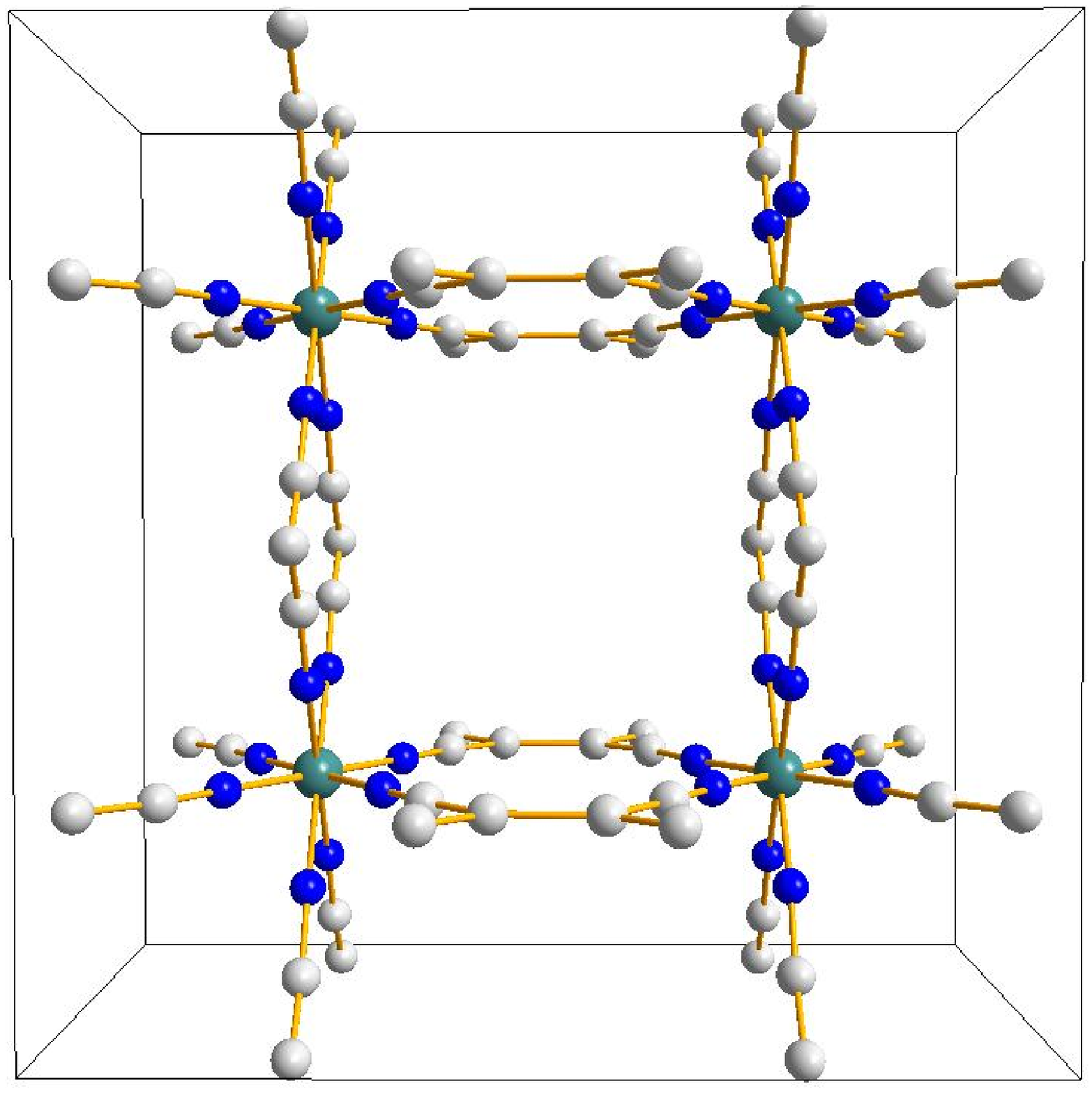} 
\includegraphics[bb = 20 20 575 805, scale=0.2]{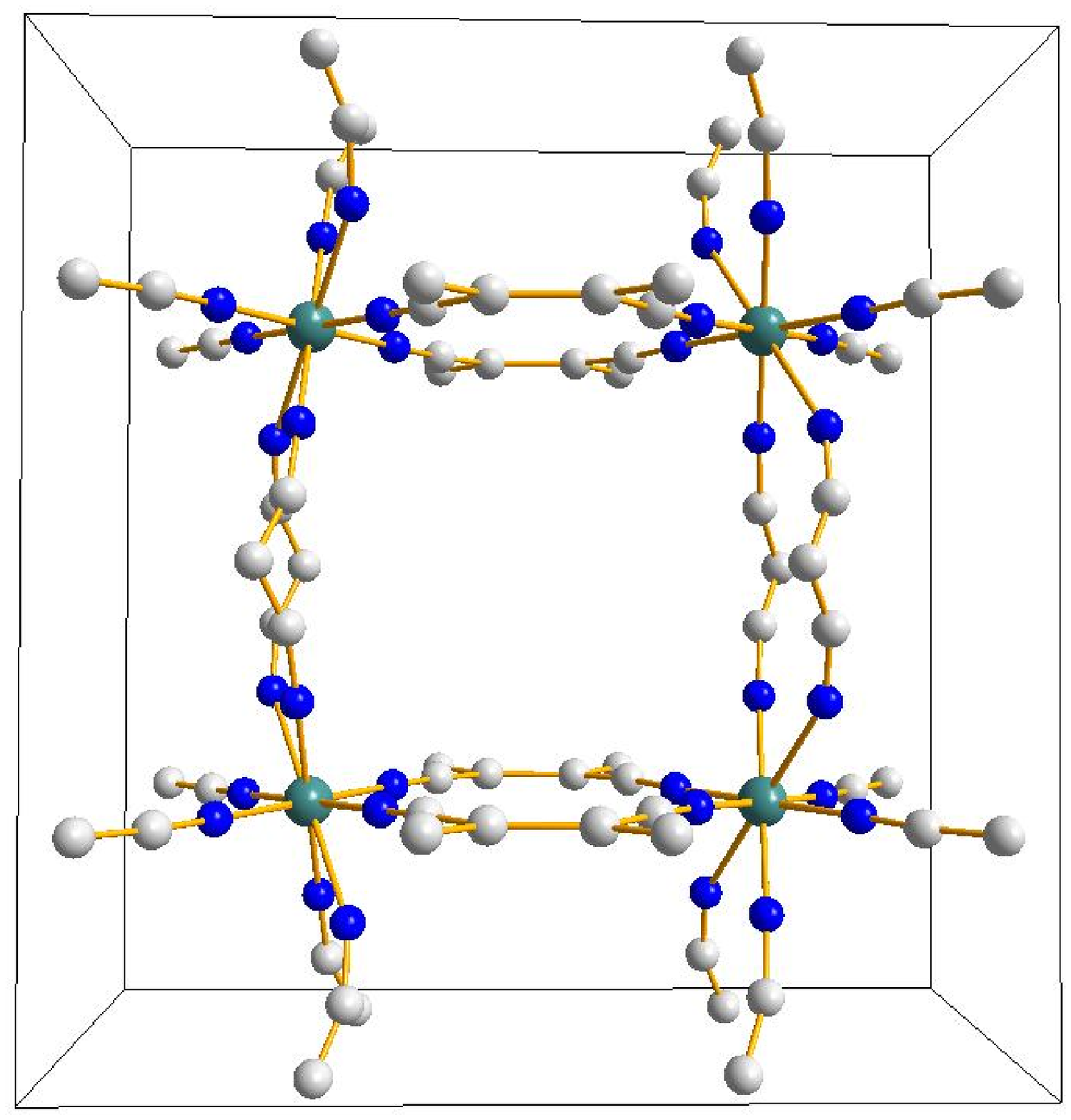} 
\includegraphics[bb = 20 20 575 805, scale=0.2]{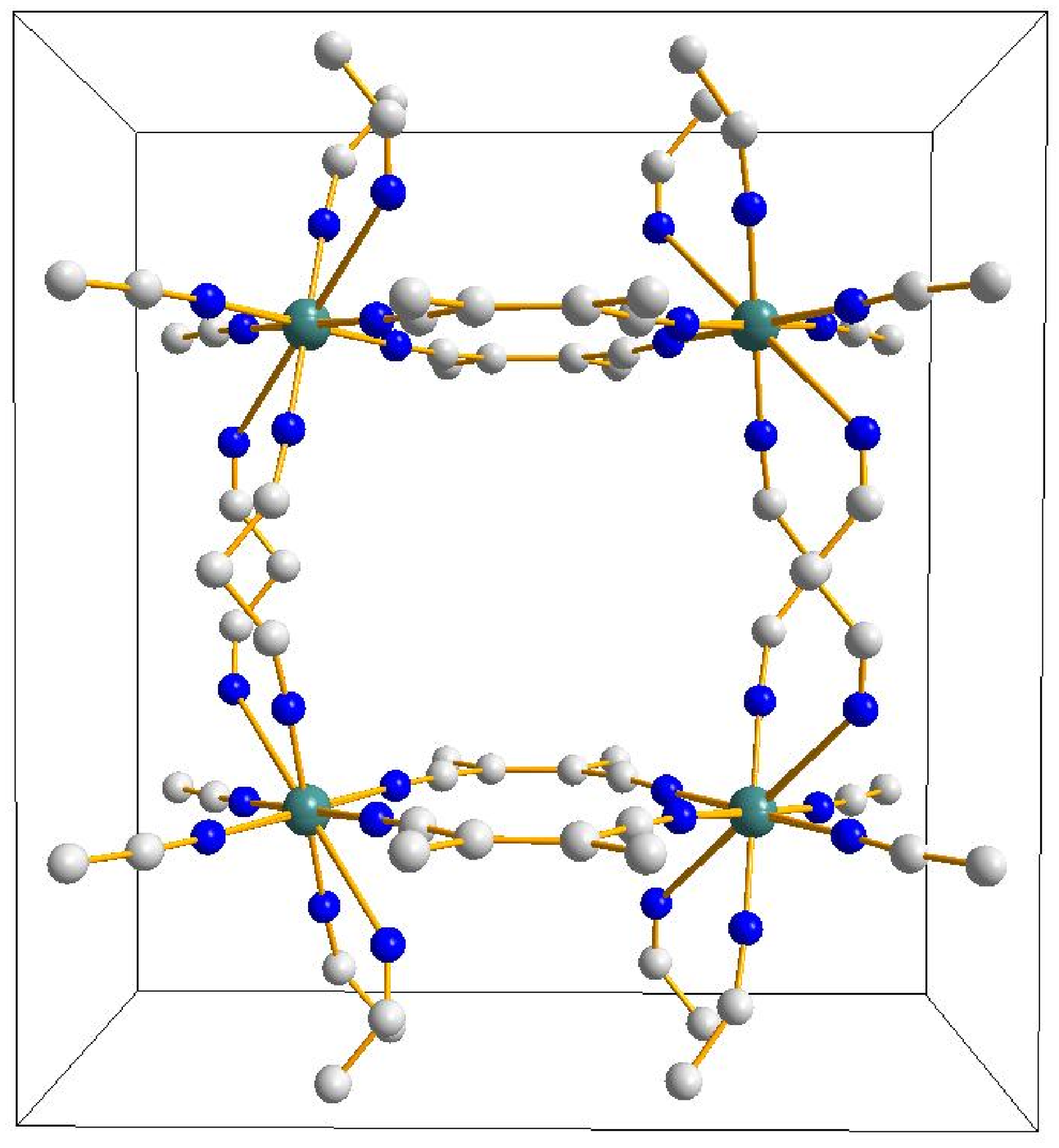} \\%
\includegraphics[bb = 20 20 575 805, scale=0.2]{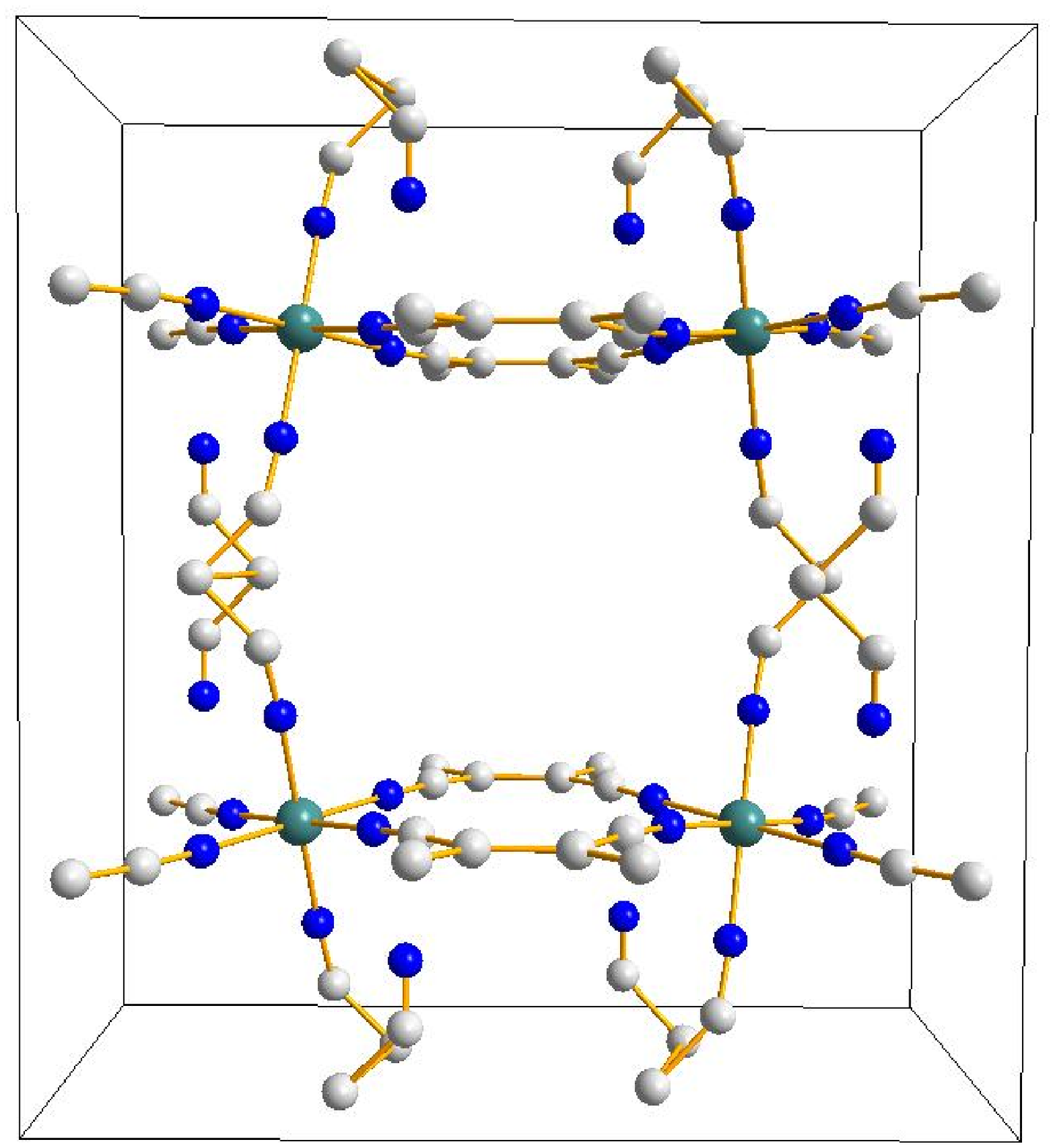} %
\includegraphics[bb = 20 20 575 805, scale=0.2]{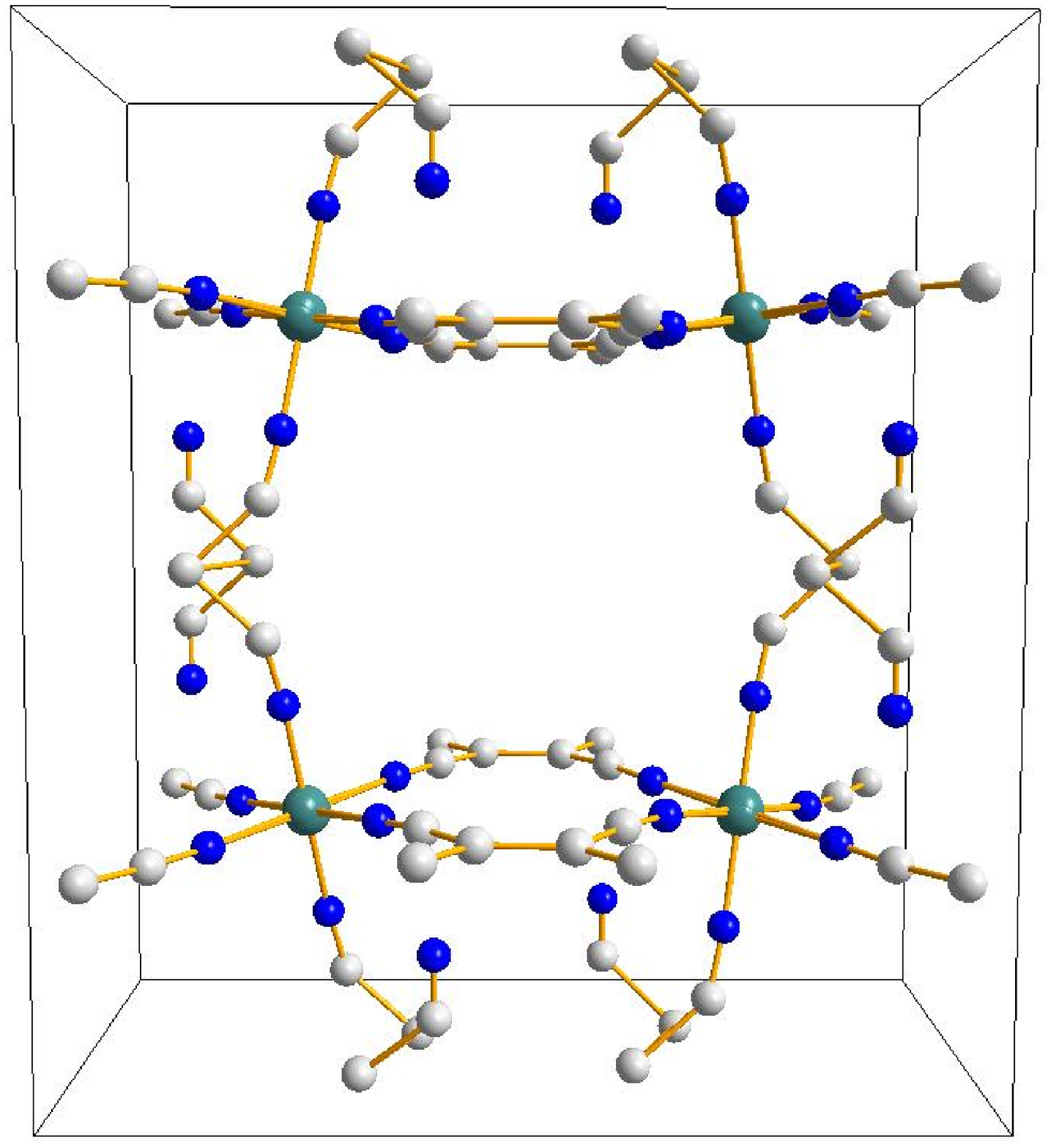} %
\includegraphics[bb = 20 20 575 805, scale=0.2]{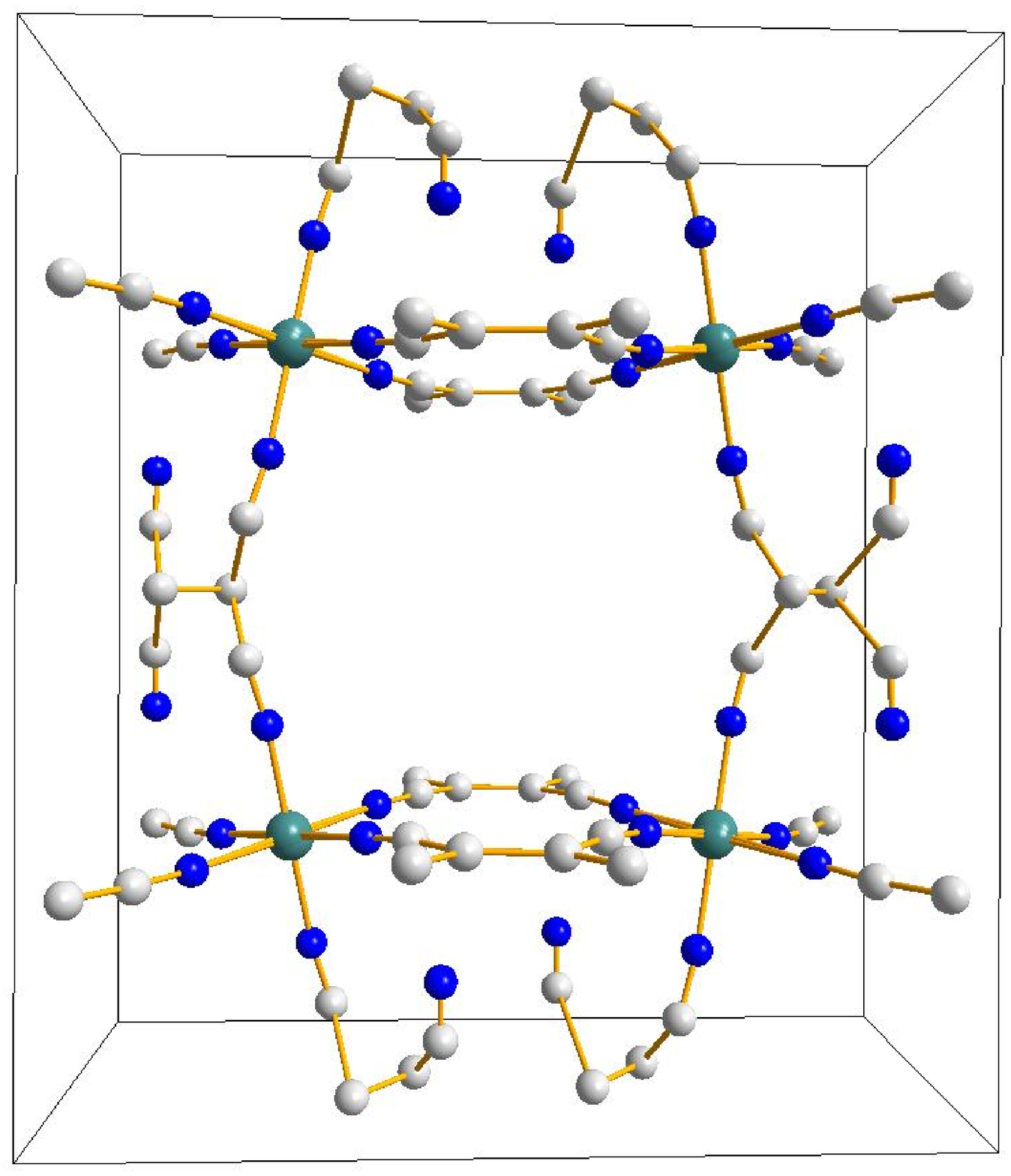}\\ %
\includegraphics[bb = 20 20 575 805, scale=0.2]{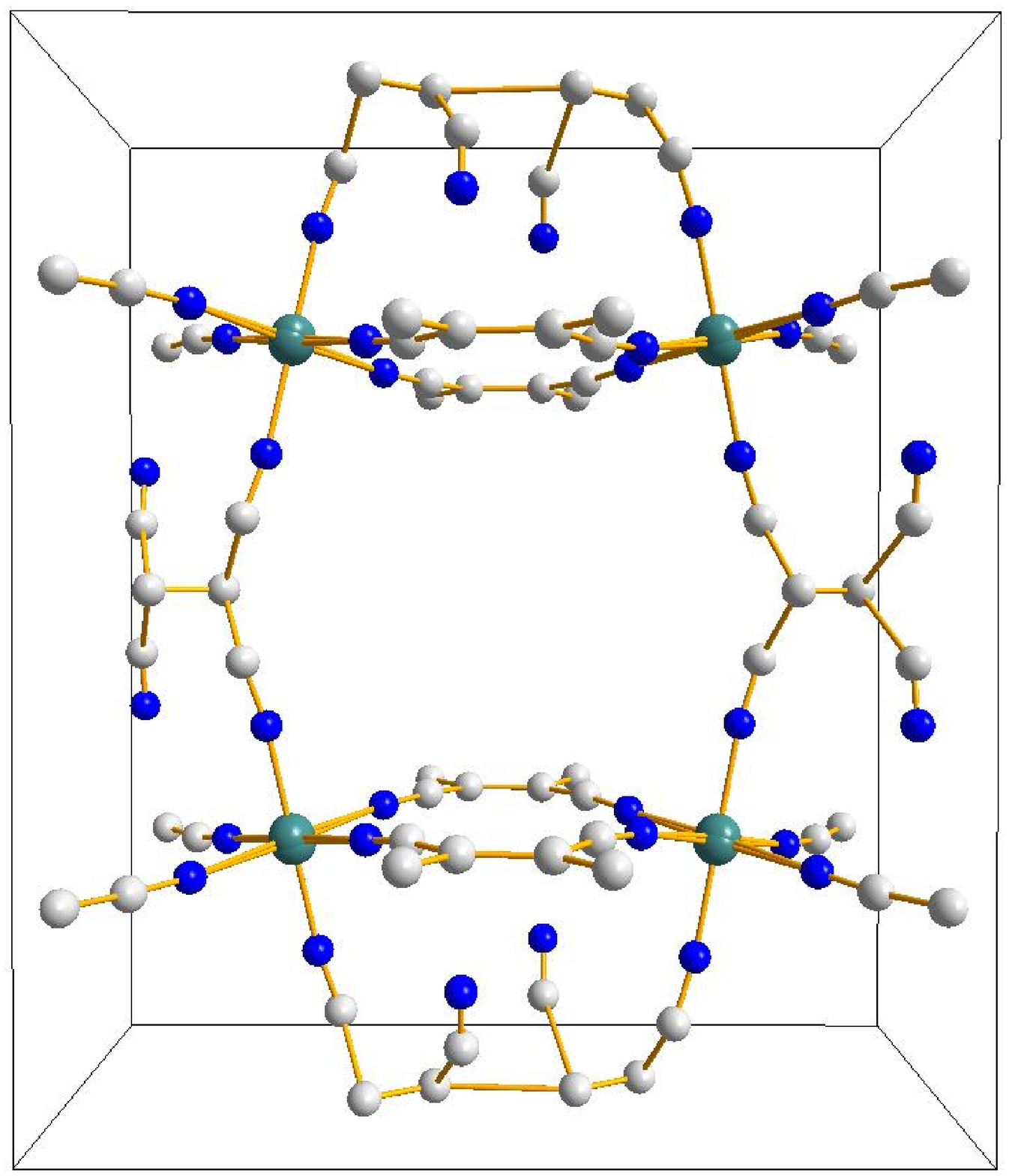} 
\includegraphics[bb = 20 20 575 805, scale=0.2]{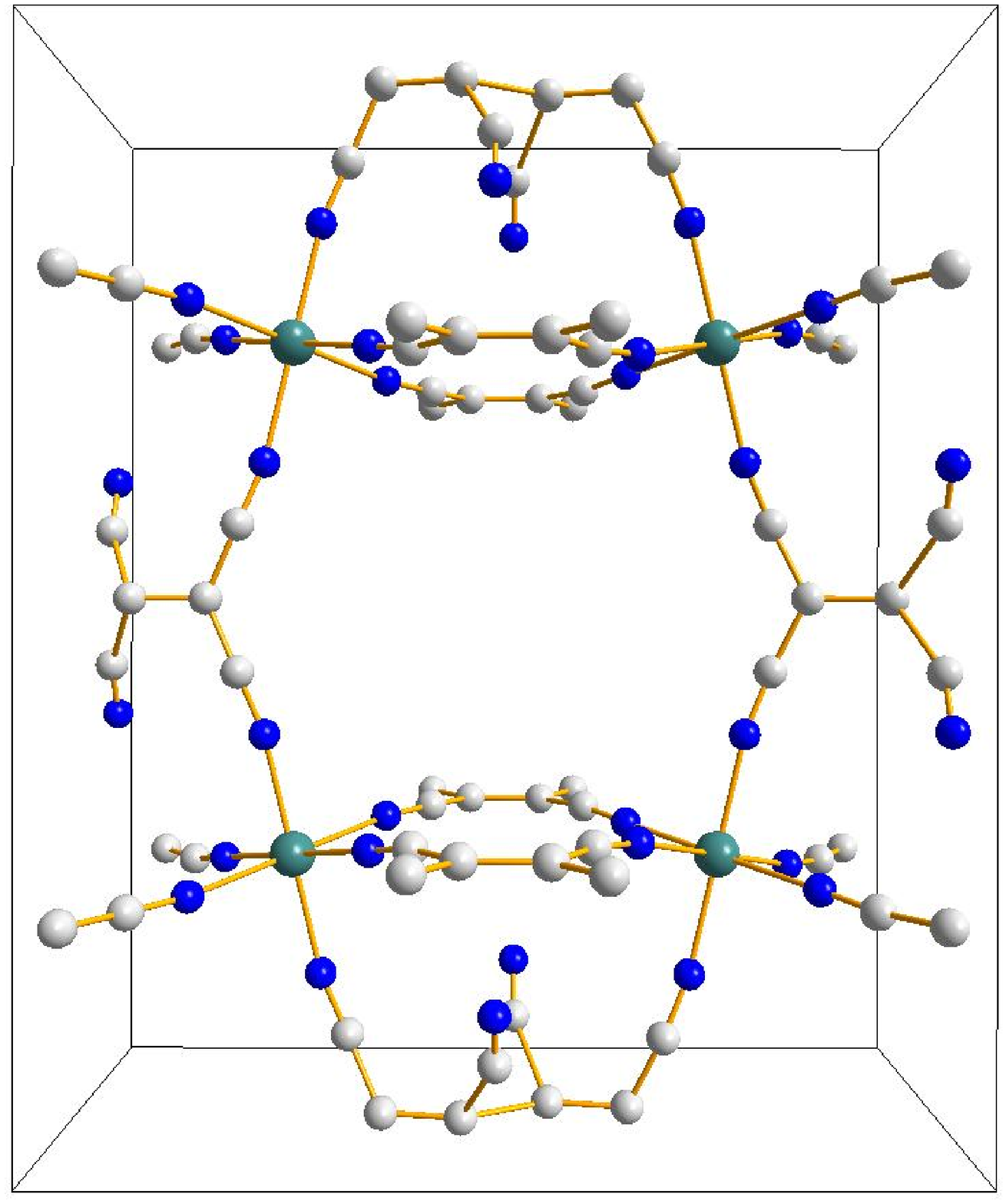} %
\includegraphics[bb = 20 20 575 805, scale=0.2]{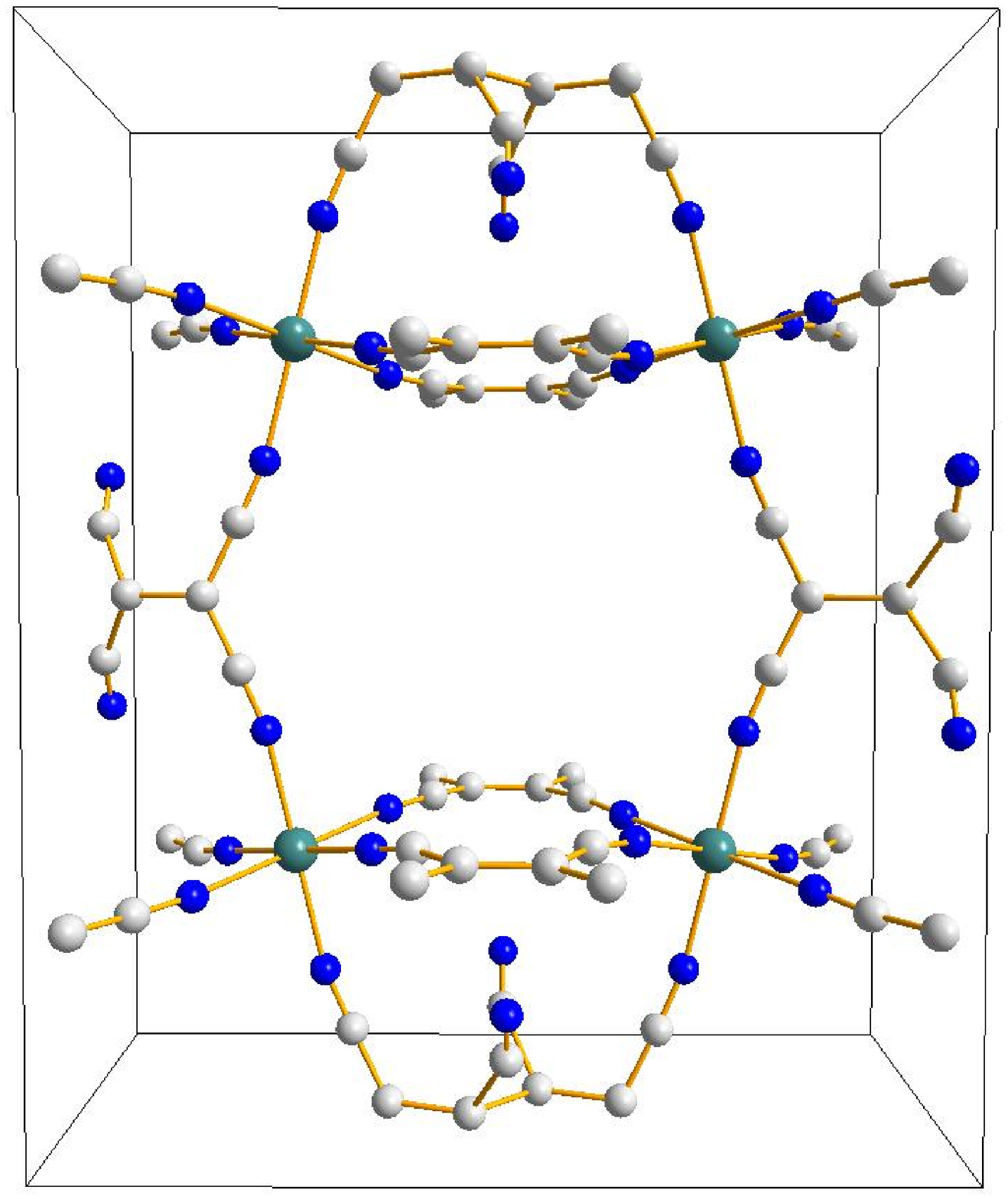} 
\center{
\caption{Schematic representation of the intermediate structures between
the quadrupled "principal" (left upper corner) and the experimental (right
lower corner) ones.}
}
\end{figure}

\begin{landscape}
\begin{figure}[tbph]
\label{vdos0-vdos10} 
\includegraphics[scale=0.8]{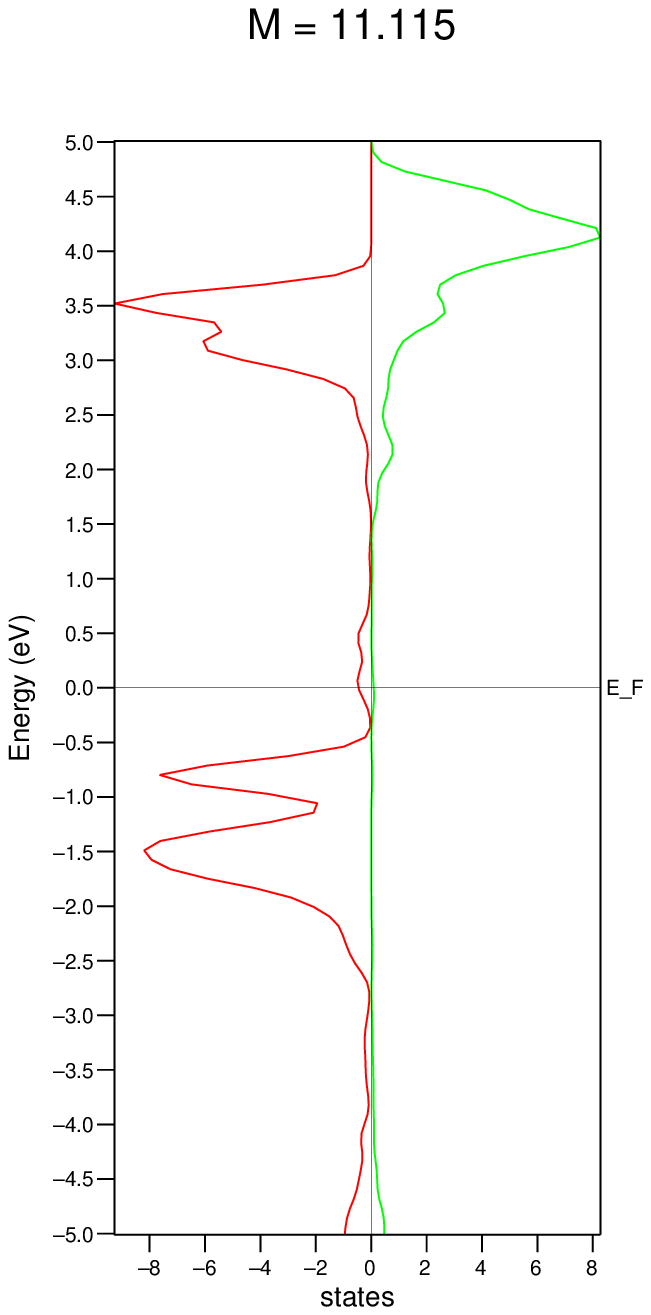} 
\includegraphics[scale=0.8]{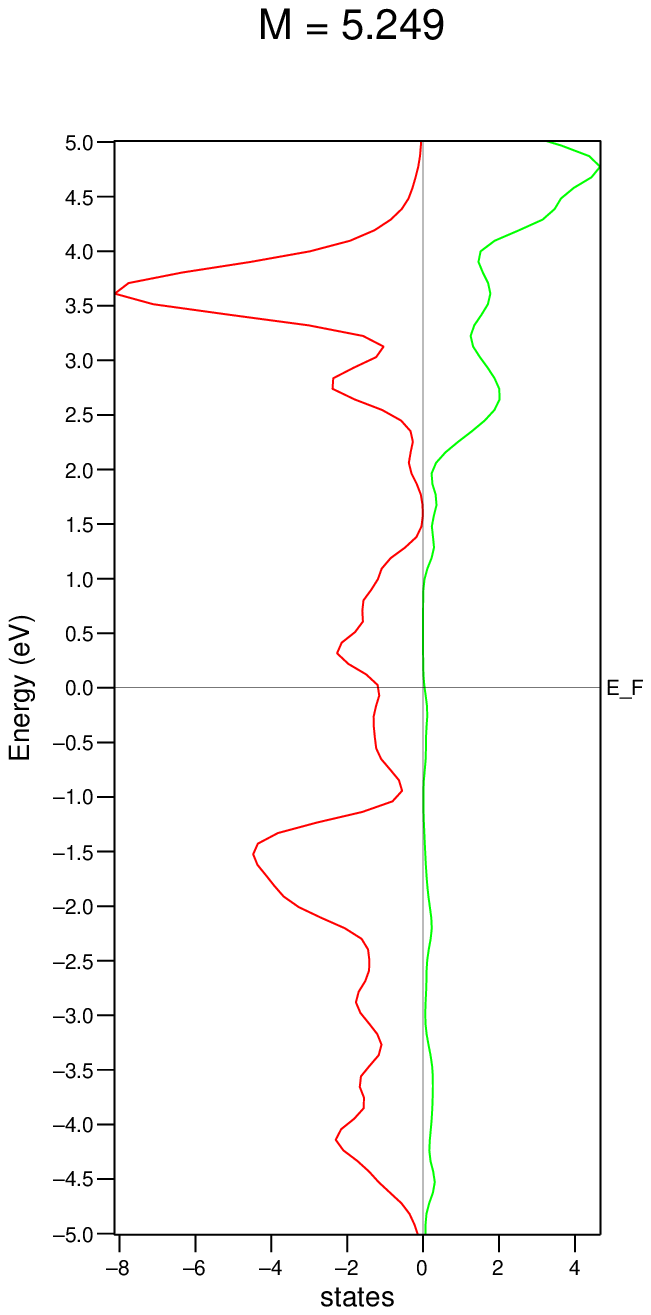} 
\includegraphics[scale=0.8]{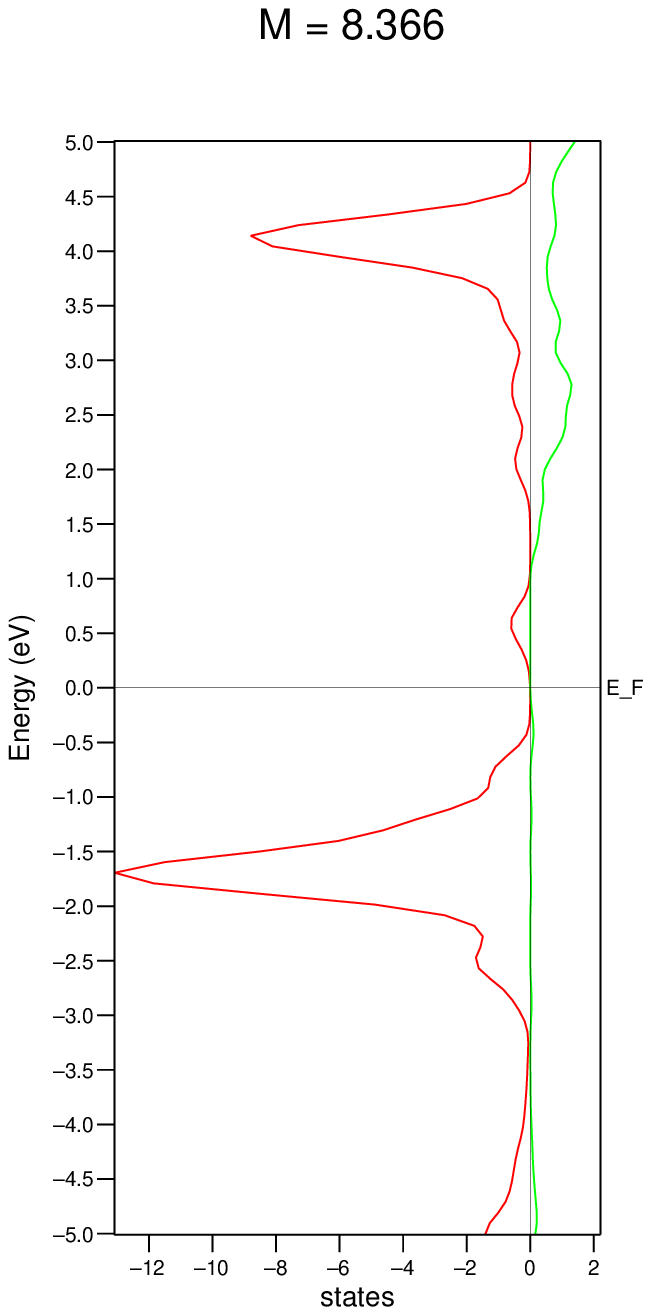} 

\caption{V-Densities of states in two spin channels for the limiting and an intermediate structures
between
the quadrupled "principal" (left) and the experimental (right) ones. Observe different scale
of the abscissa for different pictures. The magnetization in units of number of unpaired
electrons per unit cell is indicated on top of each picture. }

\end{figure}
\end{landscape}
\begin{landscape}
\begin{figure}[tbph]
\label{ndos0-ndos10} 
\includegraphics[scale=0.8]{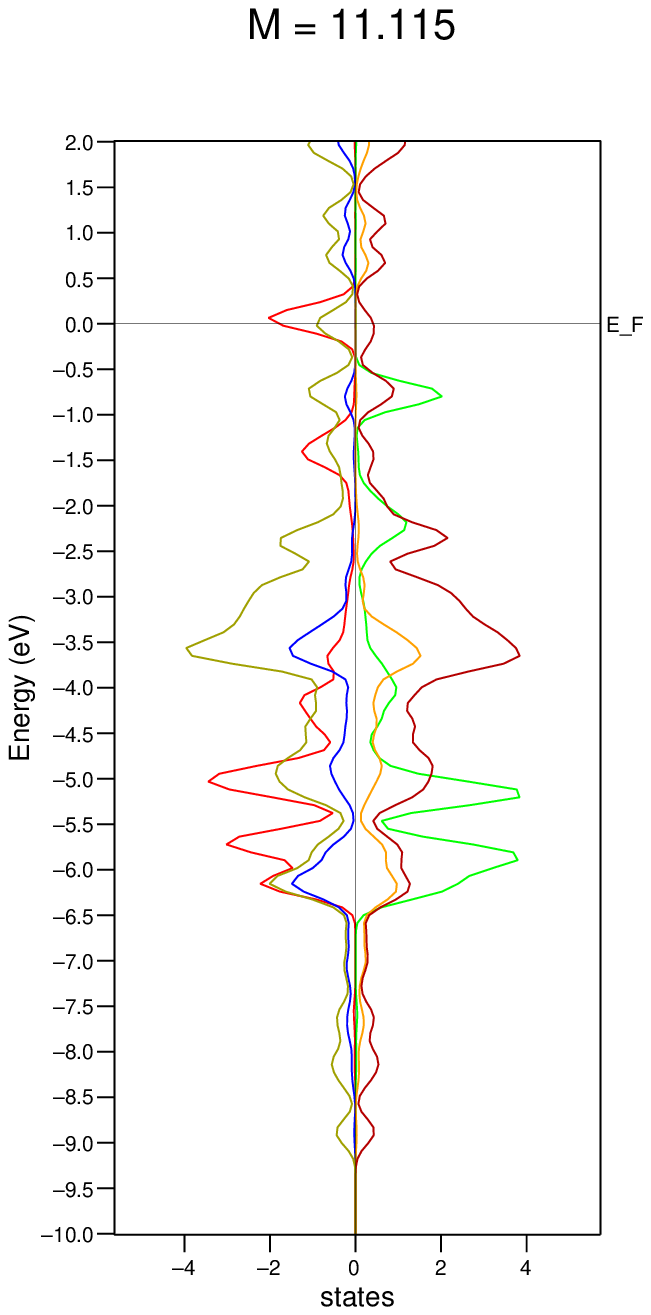} 
\includegraphics[scale=0.8]{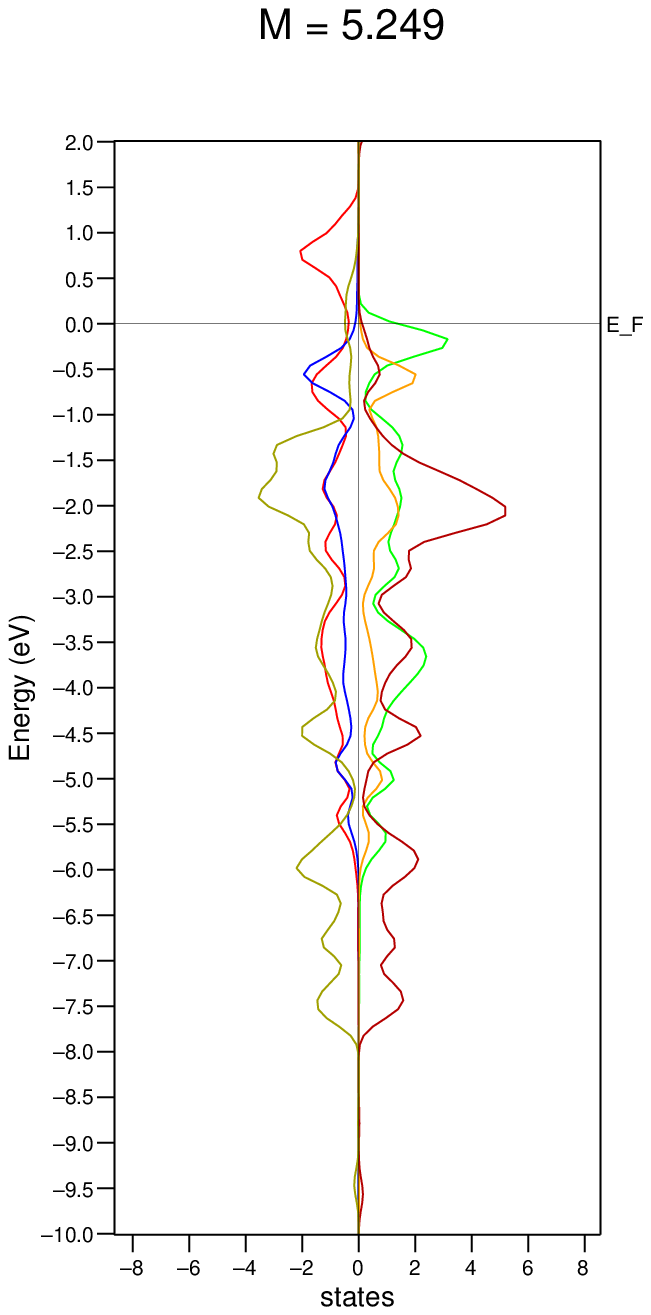} 
\includegraphics[scale=0.8]{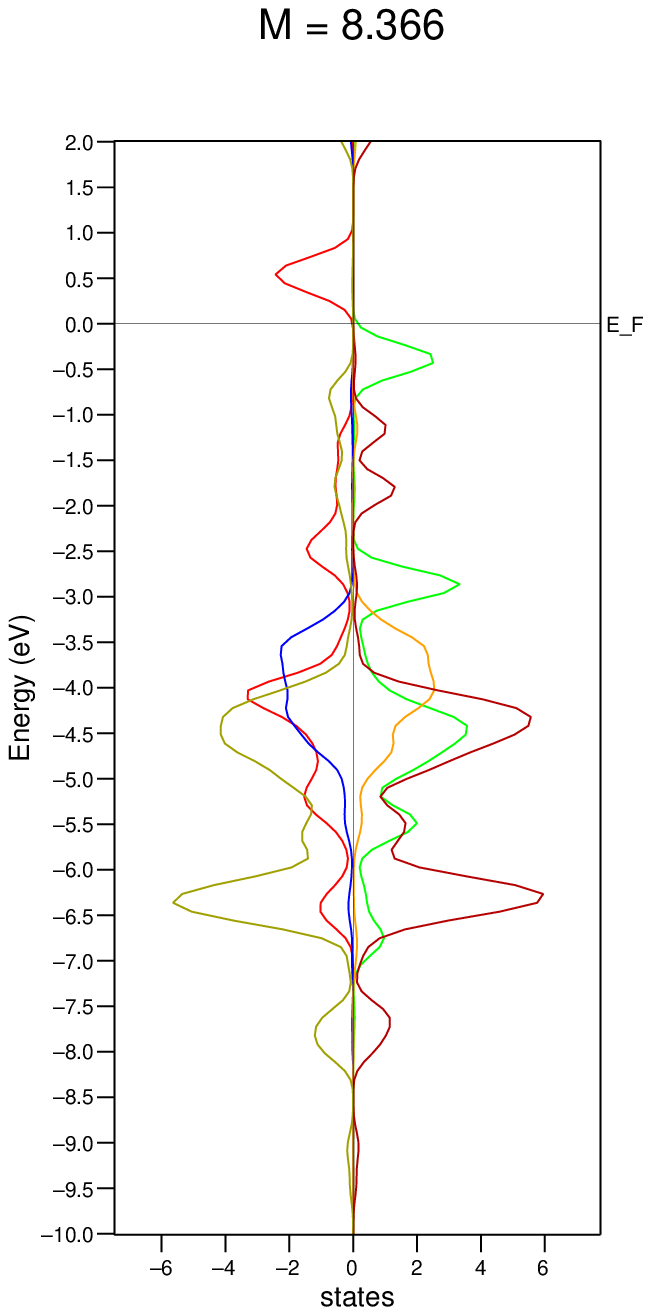} 

\caption{N-Densities of states in two spin channels for the limiting and an intermediate structures
between
the quadrupled "principal" (left) and the experimental (right) ones. Red/green 
is the DoS of the N atoms in the V-TCNE layers, blue/yellow is that of the 
"to be dangling" ones; olive green/dark red is that of those involved 
in the V-V bonding through the [TCNE]$_2^{2-}$ units. Observe different scale
of the abscissa for different pictures. The magnetization in units of number of unpaired
electrons per unit cell is indicated on top of each picture.}
\end{figure}
\end{landscape}
\begin{landscape}
\begin{figure}[tbph]
\label{cdos0-cdos10} 
\includegraphics[scale=0.8]{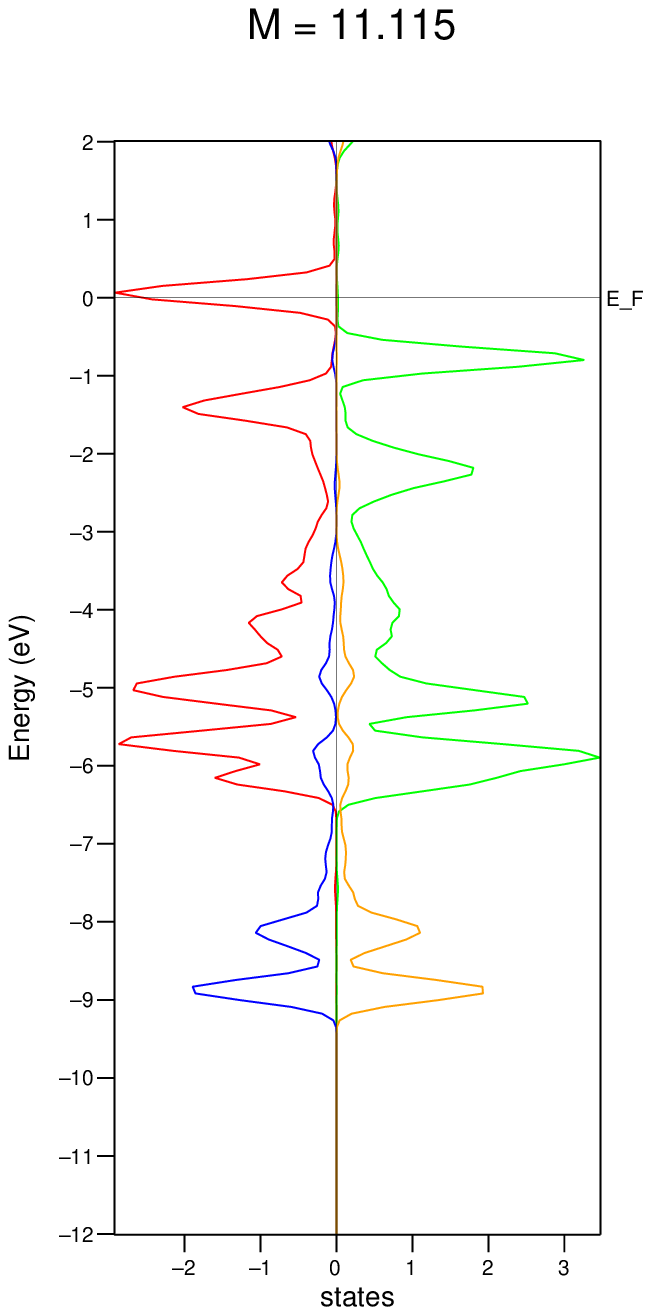} 
\includegraphics[scale=0.8]{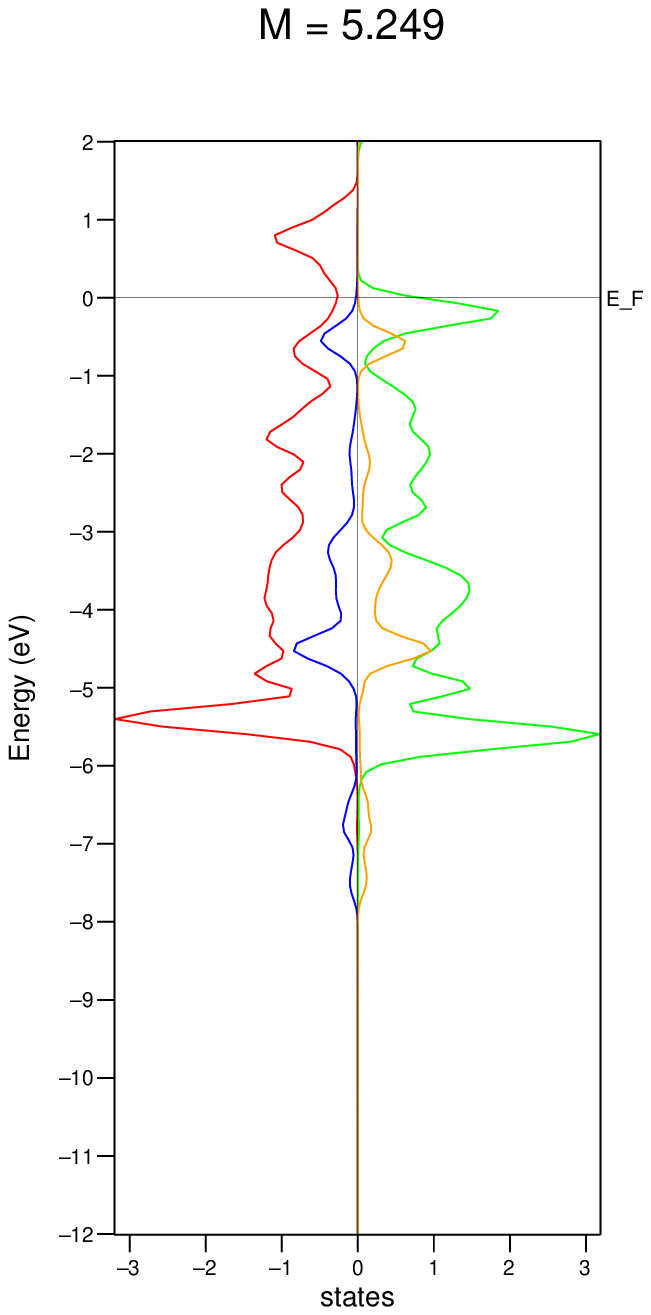} 
\includegraphics[scale=0.8]{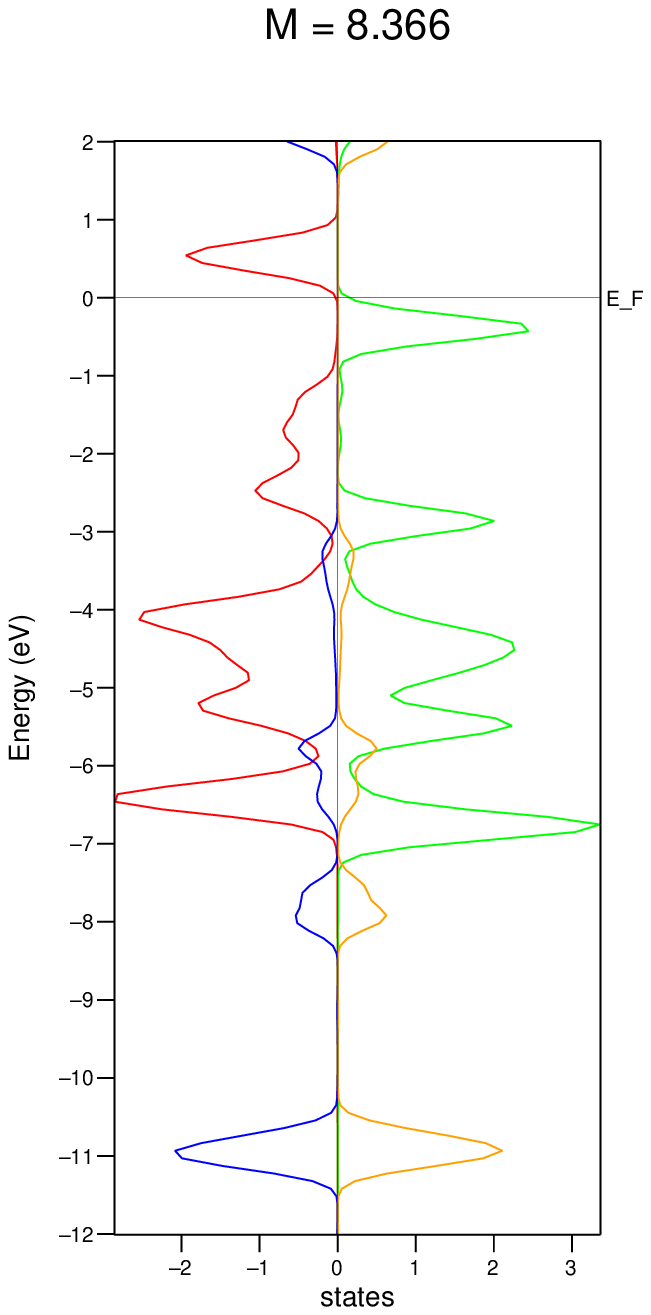} 

\caption{C-Densities of states in two spin channels for the limiting and an intermediate structures
between
the quadrupled "principal" (left) and the experimental (right) ones. Red/green 
is the DoS of the C atoms in the V-TCNE layers, blue/yellow is that of those 
forming the C-C bonds in the [TCNE]$_2^{2-}$ units. Observe different scale
of the abscissa for different pictures. The magnetization in units of number of unpaired
electrons per unit cell is indicated on top of each picture.}
\end{figure}
\end{landscape}
\begin{figure}[tbp]
\label{tcne2_DOS_10} 
\includegraphics*[bb = 25 -10 219 375]{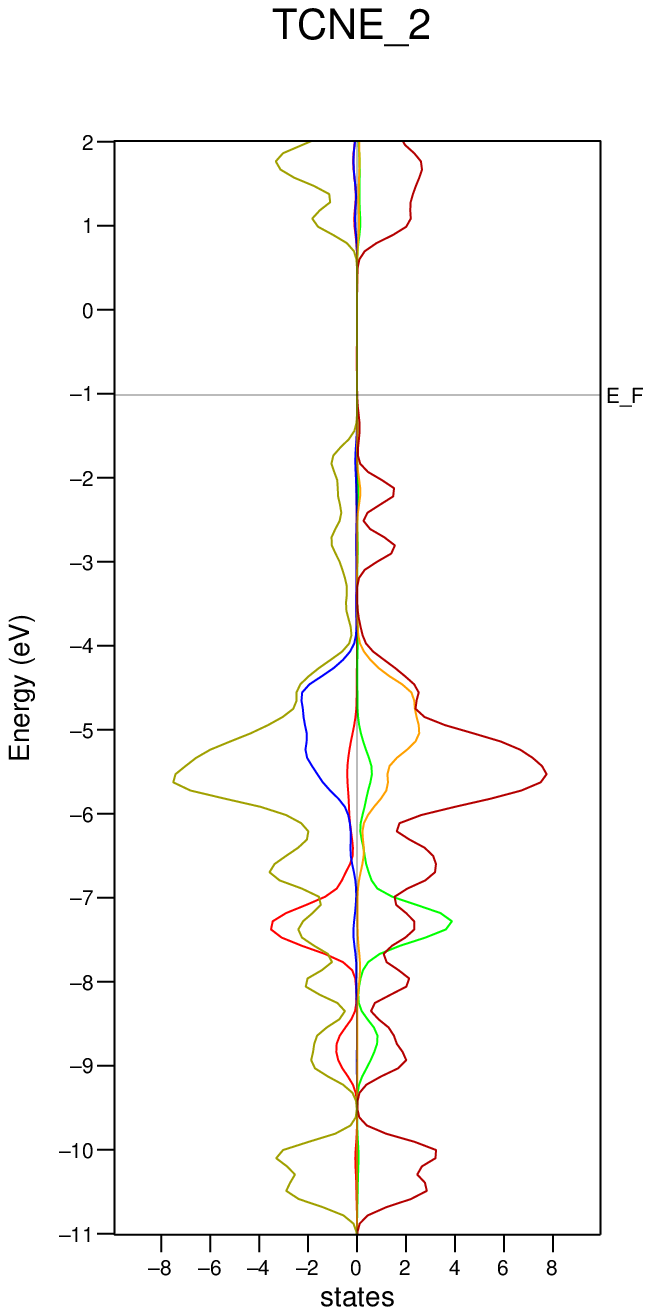}
\caption{The projection of the DoS in two spin channels to the [TCNE]$_2^{2-}$ units. 
in the "experimental" structure \cite{Her}. Red/green 
is the DoS of the N atoms bonding the V-TCNE layers; blue/yellow is that of the 
"dangling" ones; olive green/dark red is that of the C atoms.}
\end{figure}


\begin{thebibliography}{99}
\bibitem{Manriquez} J.M. Manriquez, G.T. Yee, S.R. McLean, A.J. Epstein,
J.S. Miller. Science (1991) 252, 1415.

\bibitem{Vickers1} E.B. Vickers, T.D. Selby, J.S. Miller. J. Am. Chem. Soc.
(2004) 126, 3716.

\bibitem{Vickers2} E.B. Vickers, T.D. Selby, M.S. Thorum, M.L. Taliaferro,
J.S. Miller. Inorg. Chem. (2004) 43, 6414.

\bibitem{Taliaferro} M.L. Taliaferro, M.S. Thorum, J.S. Miller. Angew. Chem.
Int. Ed. (2006) 45, 5326.

\bibitem{Pokhodnya} K.I. Pokhodnya, E.B. Vickers, M. Bonner, A.J. Epstein,
J.S. Miller. Chem. Mater. (2004) 16, 3218.

\bibitem{MillerEpstein} J.S. Miller, A.J. Epstein. Chem. Commun. (1998) 1319.

\bibitem{ZhouLongMillerEpstein} P. Zhou, S.M. Long, J.S. Miller, A.J. Epstein. 
Phys. Lett. A (1993) 181, 71.

\bibitem{Her} J.-H. Her, P.W. Stephens, K.I. Pokhodnya, M. Bonner, J.S.
Miller. Angew. Chem. Int. Ed. (2007) 46, 1521 - 1524

\bibitem{Tchougreeff-Hoffmann} A.L. Tchougr\'{e}eff, R. Hoffmann. J. Phys.
Chem. (1993) 97, 350.

\bibitem{Tch082} A.L. Tchougr\'{e}eff, R. Dronskowski. J. Comp. Chem. 
(2008) 29, 2220.

\bibitem{VASP} VASP the Guide. by G. Kresse and J. Furthm\"{u}ller, Institut
f\"{u}r Materialphysik, Universit\"{a}t Wien, (2007).

\bibitem{Goodenough} J.B. Goodenough. Magnetism and the Chemical Bond. 
Interscience-Wiley, NY (1963).

\bibitem{Kahn} O. Kahn. Molecular Magnetism. VCH (1993). 

\bibitem{Tch018} A.L. Tchougr\'{e}eff, I.A. Misurkin. Phys. Rev. B  (1992) 46, 5357.

\bibitem{Tch019} A.L. Tchougr\'{e}eff. J. Chem. Phys. (1992) 96, 6026. 

\bibitem{Vonsovsky} S.V. Vonsovskii, Magnetism. Nauka, Moscow (1971) [in
Russian]; S.V. Vonsovsky, Magnetism. Wiley, NY (1974) in two volumes.

\bibitem{Holstein-Primakoff} T. Holstein, H. Primakoff. Phys. Rev. (1940) 58, 1098. 

\bibitem{WeiQiuDu} G. Wei, R. Qiu, A. Du. Phys. Lett. A (1995) 205, 335.  

\bibitem{QiuZhang} R. Qiu, Z. Zhang. J. Phys. Cond. Matt. (2001) 13, 4165.   

\bibitem{SinghTesanovic} A. Singh, Z. Te\v{s}anovi\'c. H. Tang, G. Xiao, C.L. Chien, 
J.C. Walker. Phys. rev. Lett. (1990) 64, 2571.

\bibitem{Buschow-de-Boer} K.H.J. Buschow, F.R. de Boer. Physics of
Magnetism and Magnetic Materials. Kluwer, NY (2004).

\bibitem{Auerbach} A. Auerbach. Interacting Electrons and Quantum Magnetism.
Springer-Verlag, NY 1994.

\bibitem{PPEM2001} K.I. Pokhodnya, D. Pejakovic, A.J. Epstein,  J.S. Miller. Phys. Rev. B 
(2001) 63, 174408.

\bibitem{MillerPreprint} J. S. Miller. Polyhed. (2009) 28, 1596.

\bibitem{Katanin-Irkhin} A.A. Katanin, V.Yu. Irkhin. Usp. Fiz.
Nauk, 177 (2007) 639 - 662 [in Russian]; Physics -- Uspekhi 50 (2007) No 6.
[in English].

\bibitem{NIST} http://physics.nist.gov/PhysRefData/ASD/levels\_form.html

\bibitem{Anderson} P.W. Anderson in Marnetism. Vol. 1. G.T. Rado, H. Suhl Eds. AP, NY (1963).

\bibitem{Tch017} A.V. Soudackov, A.L. Tchougr\'{e}eff, I.A. Misurkin. Theor. Chim. Acta (1992) 83,  389.

\bibitem{Tch033} A.V. Soudackov, A.L. Tchougr\'{e}eff, I.A. Misurkin. Int. J. Quant. Chem. (1996) 57, 663.

\bibitem{Tch037} A.V. Soudackov, A.L. Tchougr\'{e}eff, I.A. Misurkin. Int. J. Quant. Chem. (1996) 58, 161. 

\bibitem{Hitchcock} P.B. Hitchcock, D.L. Hughes, G.J. Leigh, J.R. Sanders, J. De Souza, C.J. McGarry, L.F. Larkworthy J. Chem. Soc., Dalton Trans. (1994) 3683.

\bibitem{Lever} A. B. P. Lever. Inorganic Electronic Spectroscopy. 2-nd Edition.  
Elsevier, NY (1984).

\bibitem{DeFusco} G.C. De Fusco, L. Pisani, B. Montanari, and N.M. Harrison. 
Phys. Rev. B (2009) 79, 085201.
\end{thebibliography}
\end{document}